\documentclass{article}

\PassOptionsToPackage{round}{natbib}
\usepackage[preprint]{neurips_2026}

\usepackage[utf8]{inputenc}
\usepackage[T1]{fontenc}
\usepackage{hyperref}
\usepackage{url}
\usepackage{booktabs}
\usepackage{amsfonts}
\usepackage{amsmath}
\usepackage{amssymb}
\usepackage{nicefrac}
\usepackage{microtype}
\usepackage{xcolor}
\usepackage{graphicx}
\usepackage{multirow}
\usepackage{enumitem}
\usepackage[section]{placeins}

\newsavebox{\tablebox}
\newcommand{\fitbox}[1]{\sbox{\tablebox}{#1}\ifdim\wd\tablebox>\textwidth\resizebox{\textwidth}{!}{\usebox{\tablebox}}\else\usebox{\tablebox}\fi}

\hypersetup{
  colorlinks=true,
  linkcolor=blue!60!black,
  citecolor=teal!65!black,
  urlcolor=blue!65!black
}

\title{GraphScent: Calibrating Directional Relevance Signals for Bounded-Context Code Localization}

\author{%
  Ramesh Arvind Naagarajan \\
  Department of Electrical Engineering and Information Technology \\
  Chemnitz University of Technology \\
  \texttt{ramesh.naagarajan@etit.tu-chemnitz.de}
}

\begin{document}

\maketitle

\begin{abstract}
Running open-weight language models locally on a single consumer GPU requires replacing expanding multi-turn exploration histories with short, stateless code inspections. Stateless repository localization, however, hinges on a foundational question: when a frozen language model evaluates an unedited code entity from a compact view, which structural relationships make its relevance judgment informative about a nearby repair target? Studying Gemma~4 models across \(211\) real-world Python issues in \(86\) repositories, we uncover a systematic directional relevance asymmetry: zero-temperature relevance scores on enclosing scopes and upstream callers consistently separate true structural neighbors from hard lexical distractors, whereas scores on downstream callees, peer methods, historical co-change partners, and dense node embeddings remain near chance. Zero-leakage masked-scope probes and regressions controlling for lexical similarity and snippet length confirm that this asymmetry reflects functional role—callers expose symptom entrypoints delegating to defective routines, while downstream helpers appear benign alone—rather than target leakage or surface overlap. Guided by this finding, we introduce \textbf{GraphScent}, a training-free search controller that propagates out-of-repository, relation-conditioned log-likelihood ratios over runtime Typed-AST multigraphs before resolving ties via a single listwise comparison, alongside \textbf{AST Ripple}, a lightweight post-edit check for missed co-edits. Likelihood calibration prevents overconfident smaller models from chasing false-positive callers while lifting exact localization on lexically hard targets from \(12.8\%\) to \(23.4\%\) at \texttt{31B} within a single-GPU budget.\footnote{Dataset (\texttt{GraphScent-Bench}): \url{https://www.kaggle.com/datasets/rameshln/graphscent-bench-gemma4}; Executable Reproduction Notebook: \url{https://www.kaggle.com/code/rameshln/graphscent-bench-repro}; Competition Track: \url{https://www.kaggle.com/competitions/gemma-4-developer-agent-paper}.}
\end{abstract}

\section{Introduction}
\label{sec:intro}

Deploying capable software-engineering agents on local workstation hardware requires strict control over prompt length and key-value (KV) cache growth. Under 4-bit quantization-aware training (\texttt{w4a16}) \citep{gemma4_2026}, Gemma~4 \texttt{E4B}, \texttt{12B}, and \texttt{31B} occupy \(10.42\,\text{GiB}\), \(8.51\,\text{GiB}\), and \(20.28\,\text{GiB}\) of GPU memory including captured CUDA graphs—fitting on a single 24\,GB consumer accelerator whenever individual inference calls remain stateless and compact (\(<4.5\text{K}\) prompt tokens). In contrast, standard multi-turn coding scaffolds \citep{yang2024sweagent,jimenez2024swebench} append successive file reads, directory listings, and search outputs into an expanding conversation history, rapidly exhausting local VRAM while compounding early navigation mistakes \citep{zhang2026sweexplore}.

Replacing long multi-turn trajectories with bounded-context stateless probes shifts the burden of repository exploration onto an explicit external search controller. Suppose an external controller selects candidate functions or classes one at a time, renders each as a compact folded view (\(\le 60\) lines), and obtains a discrete relevance judgment \(S(u) \in \{0,1,2,3\}\) from a frozen local LLM. When \(u\) itself is \emph{not} the repair target, under what structural conditions does reading \(u\) provide diagnostic evidence about a nearby target \(v^* \in V^*\)? Prior repository-localization frameworks equip multi-turn agents with directed heterogeneous code graphs \citep{chen2025locagent,jiang2025cosil,yu2025orcaloca} or rank directed call and inheritance views via Personalized PageRank \citep{zhang2026repoatlas}, while active retrieval methods such as BAGEL \citep{kim2026bagel} and ReBOL \citep{korikov2026rebol} propagate sparse LLM relevance scores through Gaussian processes over dense passage embeddings. Yet static graph centrality ignores what the LLM discovers when reading candidate code, and symmetric kernel or graph propagation implicitly assumes that local relevance evidence decays isotropically across edges.

We investigate this question within the setting of the \emph{Gemma~4 Developer Agent Challenge and Paper Track} (\url{https://www.kaggle.com/competitions/gemma-4-developer-agent-paper}), an open benchmark for local software-engineering agents that provides \(129\) real-world Python GitHub issue-resolution tasks together with precomputed repository call graphs and 256-d node embeddings, which we extend with a second zero-overlap panel of \(102\) temporally held-out 2025--2026 issues across \(84\) additional Python repositories. Evaluating \texttt{Gemma 4 E4B}, \texttt{12B}, and \texttt{31B} across \(211\) localized tasks in \(86\) repositories (\(46{,}032\) cached live \(T=0\) node evaluations), we make four findings and contributions:
\begin{enumerate}[leftmargin=1.6em,itemsep=2pt,topsep=2pt]
  \item \textbf{Structural Graph Audit and Static Centrality Controls (Section~\ref{sec:audit}).} Auditing the benchmark's supplied Oct~13 call graphs (\(129\) public tasks), we find that \(100.0\%\) of non-self edges are stored reciprocally (\(u \to v \iff v \to u\)), \(27.3\%\) of candidate entities (\(44.3\%\) in \texttt{fastapi}) are isolated, and static Personalized PageRank (\(\mathrm{MRR}\;0.083\)) degrades lexical \texttt{BM25} (\(0.174\)). Rebuilding a runtime Typed-AST multigraph in pure Python (median \(0.17\,\text{s}\) on CPU) connects \(100.0\%\) of candidates overall and \(83.4\%\text{--}83.7\%\) strictly via non-containment call, inheritance, and override edges. Even on this repaired graph, however, static undirected and directed PageRank (\(\mathrm{MRR}\;0.268\) and \(0.205\)) trail \texttt{BM25} (\(0.357\)), showing that repository topology helps localization only when paired with live semantic judgments.
  \item \textbf{Controlled Measurement of Directional Relevance Asymmetry (Section~\ref{sec:scent}).} Evaluating ordinal \(T=0\) judgments on unedited structural neighbors against hard lexical distractors across two disjoint repository panels, we find that enclosing scopes (\texttt{parent}) and upstream callers (\texttt{called\_by}) consistently carry positive proximal signal across all three model scales, whereas downstream callees (\texttt{calls}) sit near chance despite \(2.4\times\) higher fan-out. Zero-leakage masked-parent probes, span-disjoint caller-versus-callee comparisons, and regressions controlling simultaneously for BM25 similarity and snippet length rule out target-code leakage and lexical confounds.
  \item \textbf{Calibrated Bounded-Context Search and Post-Edit Ripple Check (Sections~\ref{sec:method}--\ref{sec:eval}).} Guided by a \(5{,}760\)-run diffusion phase diagram, \textbf{GraphScent} propagates out-of-repository Laplace-smoothed log-likelihood ratios selectively along positive-scent directed relations (\(\texttt{parent}\to\texttt{child}\), \(\texttt{caller}\to\texttt{callee}\)) across \(B=24\) stateless probes, followed by a single top-6 listwise tie-break. Ablations show that out-of-repository likelihood calibration specifically stabilizes overconfident \texttt{E4B} (\(+3.8\,\text{pp}\) over uncalibrated \texttt{Fixed-Dir}), forward structural orientation beats reversed edges across all scales, and listwise tie-breaking adds \(+7.6\) to \(+12.3\,\text{pp}\) over pointwise \(\arg\max\). A supporting \(\sim 160\)-token post-edit tool (\textbf{AST Ripple}) surfaces missed reference-patch files in \(50.0\%\) of unresolved \texttt{31B} multi-file runs.
  \item \textbf{Verifiable Open Benchmark and Toolkit (Section~\ref{sec:discussion}).} We release \texttt{GraphScent-Bench} under Apache-2.0, including the standalone \texttt{graphscent} Python package, audited manifests, \(46{,}032\) cached \(T=0\) judgments, out-of-repository calibration folds, negative-result logs, and an executable Kaggle reproduction notebook.
\end{enumerate}

\section{Benchmark Panels, Structural Audit, and Static Graph Controls}
\label{sec:audit}

\subsection{Repository-Disjoint Panels, Provenance, and Exclusion Accounting}
To test whether structural relevance patterns generalize beyond the four repositories in the public competition benchmark, we evaluate two repository-disjoint Python panels totaling \(211\) entity-localized tasks across \(86\) unique repositories (\(3\text{--}2{,}140\) non-test Python files per repository, median \(45\), P90 \(237\); \(719\text{--}167{,}538\) non-test lines of code, median \(10{,}726\), P90 \(83{,}671\)).

\paragraph{Public benchmark panel (\texttt{public\_113}; \(4\) repositories, \(113\) tasks).}
The public release of the Gemma~4 Developer Agent Challenge supplies \(129\) real-world GitHub issues across \texttt{fastapi/fastapi} (\(56\) localized tasks), \texttt{Textualize/rich} (\(44\)), \texttt{psf/requests} (\(12\)), and \texttt{encode/httpx} (\(1\)). Because entity-level localization evaluates functions, methods, and classes, we apply a deterministic ground-truth mapping filter prior to evaluation: \(113\) tasks modify at least one existing non-test Python function or class, whereas \(16\) tasks are excluded upfront—\(13\) whose reference patch edits only module-top-level statements or introduces brand-new top-level symbols outside existing definitions, and \(3\) that touch only documentation, tests, or non-Python files.

\paragraph{Temporally held-out panel (\texttt{post2025\_98}; \(82\) repositories, \(98\) tasks).}
We complement the public split with \(102\) GitHub issues collected across \(84\) open-source Python repositories having \emph{zero} repository overlap with \texttt{public\_113}. Every issue/commit timestamp (\(102/102\), \(100.0\%\)) falls between \texttt{2025-01-02} and \texttt{2026-05-11} (P10 \texttt{2025-02-07}, median \texttt{2025-06-12}, P90 \texttt{2026-04-13}). Applying the identical entity-mapping rule yields \(98\) tasks across \(82\) repositories; the \(4\) excluded tasks (\texttt{montepy\_850}, \texttt{toqito\_1484}, \texttt{jsonata\_python\_14}, and \texttt{shtab\_182}) modify only module-level statements or add new symbols outside existing definitions. We refer to this split as temporally held-out 2025--2026 repositories rather than claiming guaranteed pretraining non-exposure, since public model documentation does not enumerate per-commit pretraining cutoffs.

\paragraph{Zero-context parametric recall probe.}
Before evaluating graph search, we check whether models can already recall defective file paths from parametric memory alone. Given only the repository name and issue text (zero repository files or directory trees), we prompt each model at \(T=0\) to predict up to five likely buggy file paths and stratify tasks by whether a gold filename already appears verbatim in the issue text (Table~\ref{tab:memo}). On unnamed issues in \texttt{post2025\_102} (\(n=63\)), \texttt{E4B}, \texttt{12B}, and \texttt{31B} recover any gold path on only \(3.2\%\), \(9.5\%\), and \(11.1\%\) of tasks (compared to \(8.4\%\), \(19.3\%\), and \(38.7\%\) on widely studied \texttt{public\_129} repositories), confirming that active repository search is essential.

\begin{table}[t]
  \caption{Zero-context parametric file-path recall (\texttt{Hit@5}) from repository name and issue text alone (\(T=0\)), stratified by whether a gold filename appears verbatim in the issue prompt.}
  \label{tab:memo}
  \centering
  \small
  \fitbox{\begin{tabular}{lcccccc}
    \toprule
    & \multicolumn{3}{c}{\textbf{Public Panel (\texttt{public\_129}, 4 repos)}} & \multicolumn{3}{c}{\textbf{Held-Out 2025--26 Panel (\texttt{post2025\_102}, 84 repos)}} \\
    \cmidrule(lr){2-4} \cmidrule(lr){5-7}
    \textbf{Model Scale} & All (\(129\)) & Named (\(10\)) & \textbf{Unnamed (\(119\))} & All (\(102\)) & Named (\(39\)) & \textbf{Unnamed (\(63\))} \\
    \midrule
    Gemma 4 \texttt{E4B} & \(12.4\%\) & \(60.0\%\) & \(\mathbf{8.4\%}\)  & \(31.4\%\) & \(76.9\%\) & \(\mathbf{3.2\%}\) \\
    Gemma 4 \texttt{12B} & \(24.0\%\) & \(80.0\%\) & \(\mathbf{19.3\%}\) & \(34.3\%\) & \(74.4\%\) & \(\mathbf{9.5\%}\) \\
    Gemma 4 \texttt{31B} & \(43.4\%\) & \(100.0\%\) & \(\mathbf{38.7\%}\) & \(38.2\%\) & \(82.1\%\) & \(\mathbf{11.1\%}\) \\
    \bottomrule
  \end{tabular}}
\end{table}

\begin{figure}[t]
  \centering
  \includegraphics[width=\textwidth]{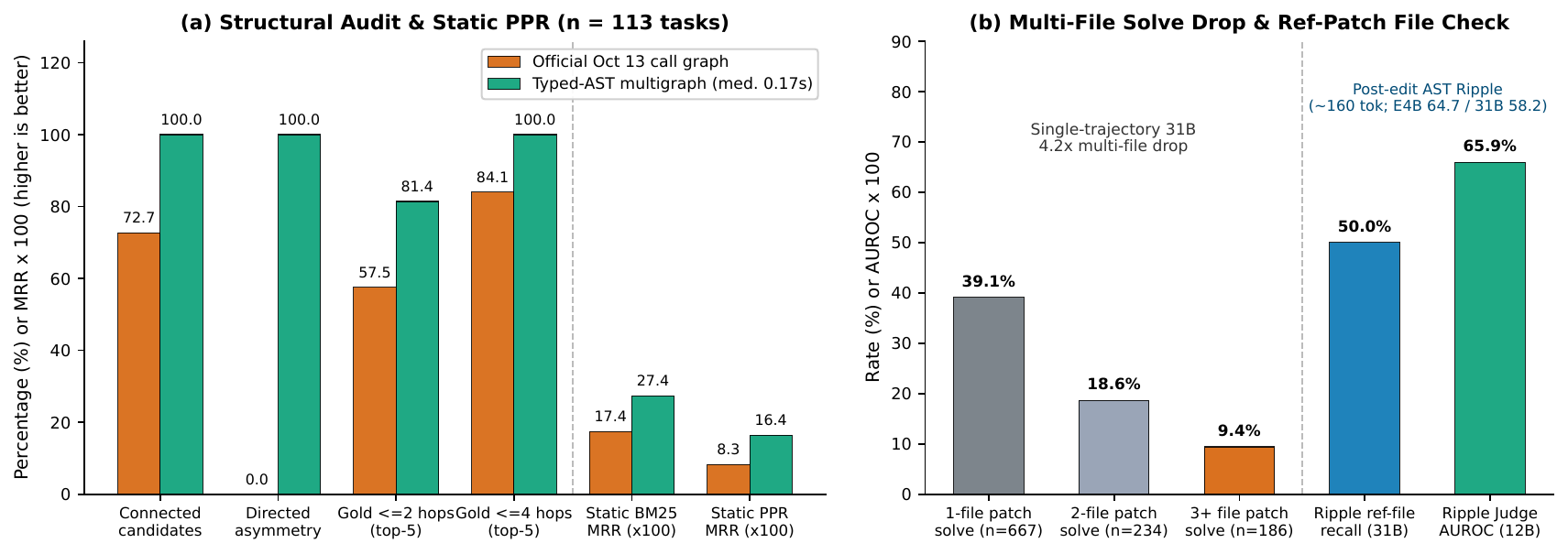}
  \caption{\textbf{Structural graph audit and multi-file co-edit gap.} \textbf{(a)} Audit of the supplied Oct~13 benchmark call graphs against our runtime Typed-AST multigraph (\(n=113\) public tasks), together with static \texttt{BM25} and Personalized PageRank (\texttt{PPR}) \texttt{MRR}. \textbf{(b)} Single-trajectory \texttt{31B} issue-resolution drop from \(1\)-file (\(39.1\%\)) to \(3+\)-file patches (\(9.4\%\)), alongside post-edit \texttt{AST Ripple} missed reference-file recall (\(50.0\%\) on \texttt{31B}) and zero-shot ripple-judge \texttt{AUROC} (\(65.9\%\) on \texttt{12B}).}
  \label{fig:audit}
\end{figure}

\subsection{Audit of Supplied Call Graphs vs.\ Runtime Typed-AST Multigraphs}
The public competition benchmark includes precomputed repository call graphs and 256-d node embeddings for all \(129\) tasks (revised Oct~13 release, which replaced an initial Sep~22 release where \(61/129\) graphs were empty). Auditing all \(460{,}018\) edge records in the Oct~13 release uncovers three structural bottlenecks that explain why naive graph baselines struggle (Figure~\ref{fig:audit}a). First, after removing \(2{,}598\) self-loops and \(96\) duplicate records, the remaining \(457{,}324\) non-self directed edges span \(228{,}662\) unordered node pairs, of which \(228{,}662/228{,}662\) (\(100.0\%\)) are stored reciprocally (\(u \to v\) and \(v \to u\)) under a single \texttt{calls} label, erasing caller--callee direction. Second, \(27.3\%\) of non-test Python candidate entities have degree zero (\(44.3\%\) in \texttt{fastapi/fastapi}, where decorator-registered route handlers lack explicit Python callers), leaving \(39/113\) tasks unreachable from issue-mentioned anchors and \(8/113\) unreachable from \texttt{BM25} top-5 seeds. Third, running Personalized PageRank (\texttt{PPR}) seeded by \texttt{BM25} over the Oct~13 graphs lowers mean reciprocal rank (\texttt{MRR}) from \(0.174\) to \(0.083\).

To recover typed directionality without external language servers or pre-built indexes, we parse non-test repository Python files on the fly using Python's standard \texttt{ast} module into a directed \textbf{Typed-AST multigraph} \(G=(V,E)\) over modules, classes, methods, and functions. Across all \(86\) benchmarked repositories on an AMD EPYC 9575F CPU, full AST construction plus \texttt{BM25} indexing takes a median of \(0.17\,\text{s}\) (P90 \(1.48\,\text{s}\), maximum \(7.94\,\text{s}\)). Each edge carries one of four structural types: lexical containment (\texttt{parent}/\texttt{child}, plus peer \texttt{sibling} edges within a class), resolved intra-repository calls (\texttt{calls}/\texttt{called\_by}; median \(677\) and \(858\) resolved call edges per repository on \texttt{public\_113} and \texttt{post2025\_98}), class inheritance (\texttt{inherits}), and same-name subclass/protocol method overrides (\texttt{mirror}).

\begin{table}[t]
  \caption{Static retrieval controls (zero LLM calls) on Typed-AST multigraphs across \texttt{public\_113}, \texttt{post2025\_98}, and \texttt{pooled\_211}. Even on fully connected directed Typed-AST graphs, static undirected and directed Personalized PageRank (\texttt{PPR}) fail to improve upon plain lexical \texttt{BM25}.}
  \label{tab:static_ppr}
  \centering
  \small
  \fitbox{\begin{tabular}{lccccccc}
    \toprule
    & \multicolumn{2}{c}{\textbf{\texttt{public\_113}}} & \multicolumn{2}{c}{\textbf{\texttt{post2025\_98}}} & \multicolumn{3}{c}{\textbf{\texttt{pooled\_211}}} \\
    \cmidrule(lr){2-3} \cmidrule(lr){4-5} \cmidrule(lr){6-8}
    \textbf{Static Ranking Method} & \texttt{R@5} & \texttt{MRR} & \texttt{R@5} & \texttt{MRR} & \texttt{R@5} & \texttt{R@20} & \texttt{MRR} \\
    \midrule
    Lexical \texttt{BM25} (folded AST views) & \(\mathbf{39.8\%}\) & \(\mathbf{0.274}\) & \(54.1\%\) & \(\mathbf{0.454}\) & \(\mathbf{46.5\%}\) & \(67.3\%\) & \(\mathbf{0.357}\) \\
    Undirected Typed-AST \texttt{PPR}        & \(25.7\%\) & \(0.167\) & \(54.1\%\) & \(0.385\) & \(38.9\%\) & \(62.6\%\) & \(0.268\) \\
    Directed Typed-AST \texttt{PPR}          & \(26.5\%\) & \(0.151\) & \(41.8\%\) & \(0.266\) & \(33.6\%\) & \(57.3\%\) & \(0.205\) \\
    \texttt{BM25} + Undirected \texttt{PPR} Fusion & \(37.2\%\) & \(0.248\) & \(\mathbf{56.1\%}\) & \(0.449\) & \(46.0\%\) & \(67.3\%\) & \(0.342\) \\
    \texttt{BM25} + Directed \texttt{PPR} Fusion   & \(38.9\%\) & \(0.267\) & \(53.1\%\) & \(0.448\) & \(45.5\%\) & \(\mathbf{67.8\%}\) & \(0.351\) \\
    \bottomrule
  \end{tabular}}
\end{table}

On \texttt{public\_113} and \texttt{post2025\_98}, Typed-AST multigraphs connect \(100.0\%\) of candidate nodes overall and connect \(83.42\%\) and \(83.74\%\) of candidates \emph{strictly through non-containment} (\texttt{calls}, \texttt{inherits}, \texttt{mirror}) edges alone, placing \(81.4\%\) (\(92/113\)) of public gold targets within \(2\) undirected hops of \texttt{BM25} top-5 (\(100.0\%\) within \(4\) hops). Yet Table~\ref{tab:static_ppr} establishes a critical control: without live LLM judgments, static PageRank over Typed-AST graphs—whether undirected or directed along \(\texttt{parent}\to\texttt{child}\) and \(\texttt{caller}\to\texttt{callee}\)—still trails plain lexical \texttt{BM25} (\(\mathrm{MRR}\;0.268\) and \(0.205\) vs.\ \(0.357\) on \texttt{pooled\_211}). Repository structure becomes informative only when coupled with semantic inspection of candidate code.

\section{Measuring Directional Relevance Asymmetry and Ruling Out Confounds}
\label{sec:scent}

\subsection{Protocol and Multi-Scale Directional Scent}
Motivated by information foraging theory \citep{pirolli1999information}, we measure how strongly an LLM's relevance judgment on an unedited neighbor cues a nearby repair target. Each AST entity \(u \in V\) is rendered into a deterministic \textbf{folded view} of at most \(60\) lines preserving its signature, decorators, docstring, control-flow/call statements, and return/raise lines (skeletonizing classes \(>60\) lines to their docstring and 1-line method headers). Given the issue description and folded view of \(u\), a frozen Gemma~4 model at \(T=0\) (\texttt{thinking=off}) emits an ordinal judgment \(S(u) \in \{0,1,2,3\}\) (\(3\): defect is directly inside \(u\); \(2\): likely involved or direct neighbor; \(1\): on execution path; \(0\): unrelated).

For any directed relation \(r\) defined relative to the unknown ground-truth edit set \(V^*\)—such as \(u \in \mathcal{N}_{\texttt{called\_by}}(V^*)\) meaning unedited \(u\) calls some \(v^* \in V^*\), versus \(u \in \mathcal{N}_{\texttt{calls}}(V^*)\) meaning some \(v^* \in V^*\) calls unedited \(u\)—we define \textbf{directional relevance scent} as the Mann--Whitney probability (\(\mathrm{AUROC}\)) that an unedited \(r\)-neighbor outranks an unedited hard lexical distractor \(u_{\texttt{far}} \in V_{\texttt{far}}\) (top-\texttt{BM25} non-gold candidates at undirected AST distance \(\ge 3\) from \(V^*\)):
\begin{equation}
  \label{eq:scent}
  \mathrm{Scent}(r) \;=\; \Pr\!\big(S(u_r) > S(u_{\texttt{far}})\big) + \tfrac{1}{2}\Pr\!\big(S(u_r) = S(u_{\texttt{far}})\big), \qquad u_r \sim \mathcal{N}_r(V^*),\;\; u_{\texttt{far}} \sim V_{\texttt{far}}.
\end{equation}
Here \(\mathrm{Scent}(r) = 0.50\) indicates zero discrimination beyond lexical similarity, \(\mathrm{Scent}(r) > 0.50\) indicates positive proximal cueing toward \(V^*\), and \(\mathrm{Scent}(r) < 0.50\) indicates suppression below lexical distractors. On \texttt{public\_113}, we also evaluate historical git co-modification (\texttt{cochange}), \(k\)-nearest neighbors in the competition's 256-d embeddings (\texttt{knear\_comp256}), and \(k\)-nearest neighbors in \texttt{Qwen3-Embedding-0.6B} (\texttt{knear\_qwen}), reporting \(1{,}000\)-resample task-cluster bootstrap \(95\%\) CIs.

\begin{table}[t]
  \caption{Multi-scale \(T=0\) directional relevance scent (\(\mathrm{AUROC}\) vs.\ \(1{,}266\) hard \texttt{BM25} \texttt{far} distractors; \(1{,}000\)-resample task-cluster bootstrap \(95\%\) CIs on \texttt{pooled\_211}, plus panel split). Non-AST channels (\texttt{knear\_qwen}, \texttt{cochange}, \texttt{knear\_comp256}) are measured on \texttt{public\_113}.}
  \label{tab:scent}
  \centering
  \small
  \setlength{\tabcolsep}{4.8pt}
  \fitbox{\begin{tabular}{llccccc}
    \toprule
    \textbf{Relation \(r\) rel.\ to \(V^*\)} & \textbf{Panel (\(n\))} & \textbf{Gemma 4 \texttt{E4B}} & \textbf{Gemma 4 \texttt{12B}} & \textbf{Gemma 4 \texttt{31B}} & \textbf{\texttt{31B} Pub} & \textbf{\texttt{31B} Post} \\
    \midrule
    \texttt{gold} (\(v \in V^*\), direct target) & Pool (\(394\)) & \(0.658\;[.63,.69]\) & \(0.738\;[.70,.78]\) & \(0.734\;[.70,.77]\) & \(0.719\) & \(0.754\) \\
    \midrule
    \texttt{parent} (enclosing scope)            & Pool (\(187\)) & \(\mathbf{0.643}\;[.60,.69]\) & \(\mathbf{0.713}\;[.67,.76]\) & \(\mathbf{0.719}\;[.67,.77]\) & \(\mathbf{0.696}\) & \(\mathbf{0.775}\) \\
    \texttt{called\_by} (upstream caller)        & Pool (\(328\)) & \(\mathbf{0.600}\;[.57,.63]\) & \(\mathbf{0.641}\;[.60,.68]\) & \(\mathbf{0.623}\;[.59,.66]\) & \(\mathbf{0.651}\) & \(\mathbf{0.587}\) \\
    \texttt{inherits} (base / subclass)          & Pool (\(22\))  & \(0.546\;[.43,.66]\) & \(0.526\;[.43,.65]\) & \(0.563\;[.47,.67]\) & \(0.652\) & \(0.524\) \\
    \texttt{mirror} (parallel override)          & Pool (\(113\)) & \(0.462\;[.39,.54]\) & \(0.523\;[.47,.59]\) & \(0.536\;[.49,.59]\) & \(0.561\) & \(0.519\) \\
    \texttt{sibling} (peer in same class)        & Pool (\(545\)) & \(0.519\;[.49,.55]\) & \(0.513\;[.48,.54]\) & \(0.506\;[.48,.53]\) & \(0.517\) & \(0.494\) \\
    \texttt{calls} (callee, \(2.4\times\) fan-out) & Pool (\(389\)) & \(0.477\;[.44,.51]\) & \(0.483\;[.45,.51]\) & \(0.487\;[.46,.51]\) & \(0.478\) & \(0.507\) \\
    \midrule
    \texttt{knear\_qwen} (Qwen3-Embed-0.6B)      & Pub (\(339\))  & \(0.556\;[.52,.59]\) & \(0.565\;[.52,.60]\) & \(0.541\;[.51,.57]\) & \(0.541\) & --- \\
    \texttt{cochange} (git commit history)       & Pub (\(220\))  & \(0.463\;[.42,.50]\) & \(0.488\;[.45,.53]\) & \(0.498\;[.46,.53]\) & \(0.498\) & --- \\
    \texttt{knear\_comp256} (Oct~13 256-d)       & Pub (\(324\))  & \(0.436\;[.40,.48]\) & \(0.455\;[.42,.49]\) & \(0.470\;[.44,.50]\) & \(0.470\) & --- \\
    \bottomrule
  \end{tabular}}
\end{table}

\begin{figure}[t]
  \centering
  \includegraphics[width=\textwidth]{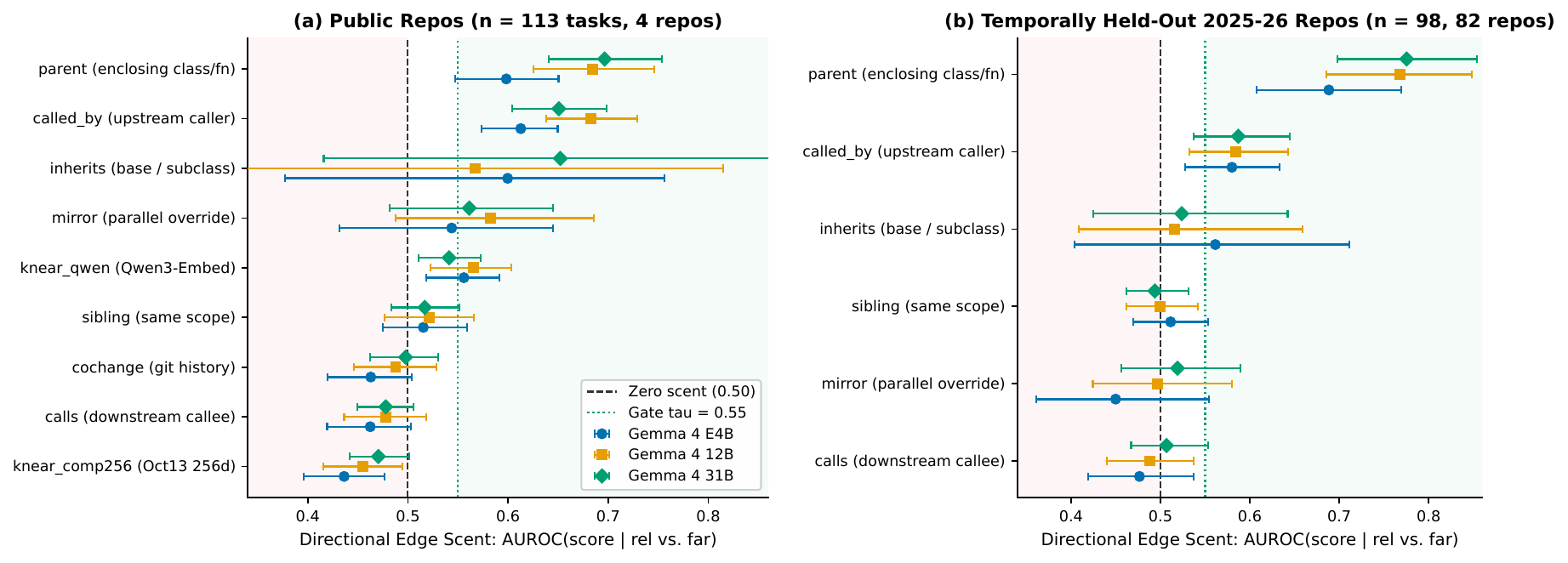}
  \caption{\textbf{Directional relevance asymmetry across model scales and disjoint repository panels.} Point estimates and \(95\%\) task-cluster bootstrap CIs for \(\mathrm{Scent}(r)\) on \textbf{(a)} \texttt{public\_113} (\(4\) repositories) and \textbf{(b)} temporally held-out \texttt{post2025\_98} (\(82\) repositories). Enclosing scopes (\texttt{parent}) and upstream callers (\texttt{called\_by}) consistently exceed the \(0.50\) distractor baseline and \(\tau=0.55\) gate across \texttt{E4B}, \texttt{12B}, and \texttt{31B}, whereas downstream callees (\texttt{calls}) remain near \(0.48\text{--}0.51\).}
  \label{fig:scent}
\end{figure}

\paragraph{Asymmetry between callers/scopes and callees.}
Table~\ref{tab:scent} and Figure~\ref{fig:scent} show a consistent directional asymmetry across all three Gemma~4 scales and both repository panels. On \texttt{pooled\_211}, \texttt{parent} scores \(0.643\), \(0.713\), and \(0.719\) across \texttt{E4B}, \texttt{12B}, and \texttt{31B}, and \texttt{called\_by} scores \(0.600\), \(0.641\), and \(0.623\) (every \(95\%\) CI strictly above \(0.56\)). By contrast, downstream callees (\texttt{calls}: \(0.477\), \(0.483\), \(0.487\)) sit at or slightly below the \(0.50\) distractor baseline while exhibiting \(2.4\times\) higher per-target multiplicity than callers. Peer methods (\texttt{sibling}: \(0.506\text{--}0.519\)), historical \texttt{cochange} (\(0.463\text{--}0.498\)), and supplied 256-d competition embeddings (\texttt{knear\_comp256}: \(0.436\text{--}0.470\)) similarly remain near chance, while \texttt{Qwen3-Embedding-0.6B} neighbors (\texttt{knear\_qwen}: \(0.541\text{--}0.565\)) carry modest signal well below \texttt{parent} and \texttt{called\_by}.

\paragraph{Scale-dependent distractor overconfidence.}
Score histograms on \texttt{far} distractors (\texttt{public\_113} / \texttt{post2025\_98}) expose a striking calibration difference across model sizes: \texttt{E4B} assigns the top score \(S=3\) to \(50.9\% / 26.5\%\) of distractors, compared to \(17.4\% / 9.9\%\) for \texttt{12B} and \(9.6\% / 7.5\%\) for \texttt{31B}. Even under this high false-positive rate, \texttt{E4B} separates \texttt{called\_by} from \texttt{far} (\(\mathrm{AUROC}\;0.600\)) because it assigns \(S=0\) far less often to true callers than to distractors (\(6.6\%\) vs.\ \(28.2\%\) on \texttt{public\_113}), motivating full ordinal likelihood calibration rather than raw-score propagation.

\subsection{Zero-Leakage Masked-Parent and Lexical/Length Confound Controls}
Before attributing positive \texttt{parent} and \texttt{called\_by} scent to structural role, we test two confounds that could inflate neighbor scores (Table~\ref{tab:confounds}).

\paragraph{Zero-leakage control for \texttt{parent} views.}
When an enclosing class has \(\le 60\) lines, its folded view includes short method bodies and could expose lines of an edited child method \(v^* \in V^*\). We re-scored all \(187\) \texttt{parent} nodes across \texttt{E4B}, \texttt{12B}, and \texttt{31B} under two strict zero-leakage transformations: \texttt{parent\_gold\_masked}, replacing every line of any gold child method with \texttt{\# [method body omitted]}, and \texttt{parent\_header\_doc}, stripping \emph{all} method bodies to keep only the class signature, class docstring, and 1-line non-gold method headers. Table~\ref{tab:confounds} shows zero degradation: \texttt{parent\_gold\_masked} achieves \(0.654\), \(0.700\), and \(0.700\), \texttt{parent\_header\_doc} achieves \(0.654\), \(0.716\), and \(0.728\), and already-skeletonized long classes (\(>60\) lines, \(n=114\)) achieve \(0.678\), \(0.785\), and \(0.811\). Enclosing-scope signal comes from class-level responsibilities and docstrings rather than leaked target lines.

\paragraph{Span-disjoint, lexical, and length controls for \texttt{called\_by} vs.\ \texttt{calls}.}
First, \texttt{far} distractors have \(2.1\times\) \emph{higher} mean normalized \texttt{BM25} similarity (\(0.634\)) than \texttt{called\_by} (\(0.308\)) and nearly identical folded length (\(38.5\) vs.\ \(38.7\) lines), so caller scent overcomes a substantial lexical disadvantage relative to \texttt{far}. Second, restricting strictly to zero-span-overlap nodes (\(297\) callers vs.\ \(372\) callees), direct discrimination \(\mathrm{AUROC}(\texttt{called\_by}\;\text{vs.}\;\texttt{calls})\) is \(0.630\;[.59,.67]\) (\texttt{E4B}), \(0.652\;[.61,.69]\) (\texttt{12B}), and \(0.629\;[.59,.67]\) (\texttt{31B}). Third, OLS regressions of \(S(u)\) on a caller indicator (\(\texttt{called\_by}=1,\;\texttt{calls}=0\)) controlling simultaneously for normalized \texttt{BM25} similarity and snippet length (\(n=669\)) yield positive caller slopes at every scale: \(\hat{\beta}_{\text{caller}} = +0.471\;[+.27,+.68]\) (\texttt{E4B}), \(+0.344\;[+.14,+.55]\) (\texttt{12B}), and \(+0.221\;[+.05,+.40]\) (\texttt{31B}), while high-\texttt{BM25} (\(\ge 0.45\)) \texttt{31B} \(\mathrm{AUROC}(\texttt{called\_by}\;\text{vs.}\;\texttt{far})\) reaches \(0.756\;[.69,.82]\). These observations are consistent with callers exposing user-facing symptom entrypoints that delegate work to defective callees, whereas generic downstream utilities called by \(v^*\) look self-contained and benign in isolation.

\begin{table}[t]
  \caption{Zero-leakage \texttt{parent} controls and span-disjoint, lexical, and length controls for \texttt{called\_by} vs.\ \texttt{calls} on \texttt{pooled\_211} (\(95\%\) task-cluster bootstrap CIs in brackets).}
  \label{tab:confounds}
  \centering
  \small
  \fitbox{\begin{tabular}{lcccc}
    \toprule
    \textbf{Control Experiment (\texttt{pooled\_211})} & \textbf{Sample \(n\)} & \textbf{Gemma 4 \texttt{E4B}} & \textbf{Gemma 4 \texttt{12B}} & \textbf{Gemma 4 \texttt{31B}} \\
    \midrule
    Original \texttt{parent} vs.\ \texttt{far}                        & \(187\) vs.\ \(1266\) & \(0.643\;[.60,.69]\) & \(0.713\;[.67,.76]\) & \(0.719\;[.67,.77]\) \\
    \texttt{parent\_gold\_masked} (0 gold child lines) vs.\ \texttt{far} & \(187\) vs.\ \(1266\) & \(\mathbf{0.654}\;[.61,.70]\) & \(\mathbf{0.700}\;[.66,.75]\) & \(\mathbf{0.700}\;[.66,.75]\) \\
    \texttt{parent\_header\_doc} (hdr + doc + non-gold sigs)          & \(187\) vs.\ \(1266\) & \(\mathbf{0.654}\;[.61,.70]\) & \(\mathbf{0.716}\;[.67,.77]\) & \(\mathbf{0.728}\;[.68,.77]\) \\
    Skeletonized \texttt{parent} (\(>60\) lines) vs.\ \texttt{far}    & \(114\) vs.\ \(1266\) & \(0.678\;[.63,.73]\) & \(0.785\;[.73,.84]\) & \(0.811\;[.77,.87]\) \\
    \midrule
    Span-disjoint \texttt{called\_by} vs.\ \texttt{far}               & \(297\) vs.\ \(1266\) & \(0.592\;[.56,.63]\) & \(0.652\;[.61,.69]\) & \(0.638\;[.60,.68]\) \\
    Span-disjoint \texttt{called\_by} vs.\ span-disjoint \texttt{calls} & \(297\) vs.\ \(372\) & \(\mathbf{0.630}\;[.59,.67]\) & \(\mathbf{0.652}\;[.61,.69]\) & \(\mathbf{0.629}\;[.59,.67]\) \\
    High-\texttt{BM25} (\(\ge 0.45\)) \texttt{called\_by} vs.\ \texttt{far} & \(58\) vs.\ \(964\) & \(0.650\;[.60,.70]\) & \(0.756\;[.70,.81]\) & \(0.756\;[.69,.82]\) \\
    OLS slope \(\hat{\beta}_{\text{caller}}\) (ctrl.\ \texttt{BM25} + lines) & \(n = 669\) & \(\mathbf{+0.471}\;[+.27,+.68]\) & \(\mathbf{+0.344}\;[+.14,+.55]\) & \(\mathbf{+0.221}\;[+.05,+.40]\) \\
    \bottomrule
  \end{tabular}}
\end{table}

\section{Method: Calibrated Bounded-Context Search and AST Ripple}
\label{sec:method}

\begin{figure}[t]
  \centering
  \includegraphics[width=\textwidth]{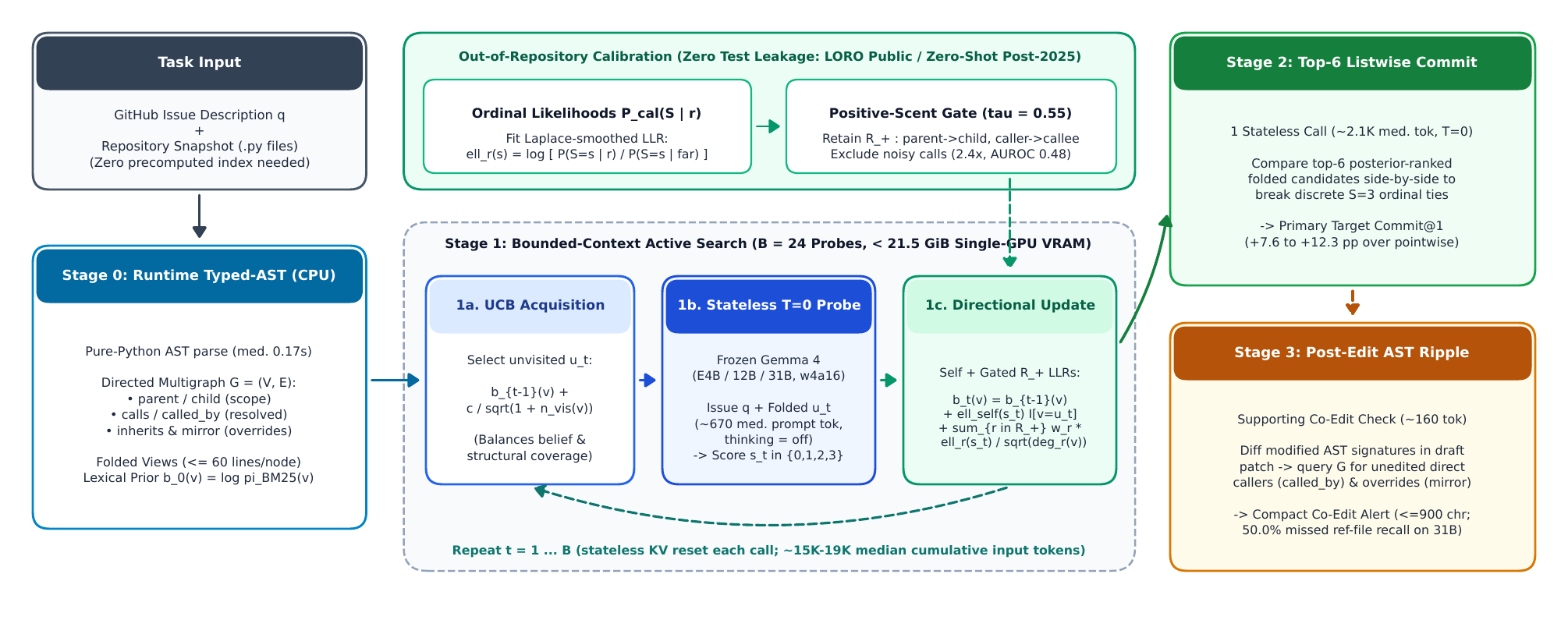}
  \caption{\textbf{System architecture of GraphScent and AST Ripple.} Stage~0 parses repository Python files into a directed Typed-AST multigraph (median \(0.17\,\text{s}\) on CPU) and initializes lexical \texttt{BM25} log-beliefs. Offline out-of-repository calibration fits ordinal log-likelihood ratios \(\ell_r(s)\) and gates propagation to \(\mathcal{R}_+ = \{r : \mathrm{Scent}_{\text{cal}}(r) \ge 0.55\}\) (\(\texttt{parent}\to\texttt{child}\), \(\texttt{caller}\to\texttt{callee}\)). Stage~1 executes \(B=24\) stateless \(T=0\) folded-view probes (\(<21.5\,\text{GiB}\) single-GPU VRAM), Stage~2 resolves discrete score ties via a single top-6 listwise commitment call, and Stage~3 runs post-edit \texttt{AST Ripple} (\(\sim 160\) tokens) to surface unedited callers and overrides.}
  \label{fig:arch}
\end{figure}

\subsection{When Does Graph Propagation Help? Linear-Diffusion Phase Diagram}
To understand when neighbor propagation improves bounded-budget (\(B=24\)) target discovery, we simulated active search under linear score diffusion across \(5{,}760\) rollouts spanning a \(12 \times 12\) grid of direct judge discrimination \(\mathrm{AUROC}(\texttt{gold}\;\text{vs.}\;\texttt{far}) \in [0.58, 0.99]\) and neighbor edge scent \(\mathrm{Scent}(r) \in [0.44, 0.80]\) on \(40\) public repository graphs (Figure~\ref{fig:results}a). Within the empirical discrimination range of open-weight judges (\(0.65\text{--}0.74\)), propagating scores along low-scent edges (\(\mathrm{Scent}(r) \le 0.52\)) lowers multi-hop \texttt{Found@24} by up to \(-15\,\text{pp}\) because false-positive scores diffuse across \(2.4\times\) high-fan-out \texttt{calls} edges and displace genuine lexical candidates. Net discovery gain turns positive only when traversal restricts to directed relations with \(\mathrm{Scent}(r) \ge 0.55\).

\subsection{GraphScent Active Search Controller and Listwise Commitment}
Guided by this threshold, \textbf{GraphScent} executes three lightweight stages (Figure~\ref{fig:arch}):

\paragraph{Stage 0: Lexical log-prior.}
Given issue description \(q\) and runtime Typed-AST multigraph \(G=(V,E)\), we compute normalized \texttt{BM25} similarity over folded entity views to define prior \(\pi(v) \propto \exp(\alpha\,\widetilde{\mathrm{BM25}}(q,v))\) and initialize log-belief \(b_0(v) = \log \pi(v)\).

\paragraph{Stage 1: Out-of-repository calibrated directional propagation.}
Using strictly out-of-repository calibration data—leave-one-repository-out cross-validation across the \(4\) \texttt{public\_113} repositories, and a single frozen \texttt{public\_113} table transferred zero-shot to all \(82\) \texttt{post2025\_98} repositories—we estimate Laplace-smoothed ordinal likelihoods \(\widehat{P}_{\text{cal}}(S=s \mid \text{rel}=r)\) for \(s \in \{0,1,2,3\}\) and define relation-conditioned log-likelihood ratios (\texttt{LLR}s) and gated relation set \(\mathcal{R}_+\):
\begin{equation}
  \label{eq:llr}
  \ell_r(s) \;=\; \log \frac{\widehat{P}_{\text{cal}}(S=s \mid \text{rel}=r)}{\widehat{P}_{\text{cal}}(S=s \mid \text{rel}=\texttt{far})}, \qquad \mathcal{R}_+ \;=\; \big\{\,r \;:\; \mathrm{Scent}_{\text{cal}}(r) \ge \tau\,\big\},
\end{equation}
with \(\tau = 0.55\) (which selects \texttt{parent} and \texttt{called\_by} across every fold, plus \texttt{inherits}/\texttt{mirror} where supported, while rejecting \texttt{calls} and \texttt{sibling} across \(\tau \in \{0.52, 0.55, 0.58\}\); Appendix~\ref{app:calibration}). Propagation follows the diagnostic orientation: when probe \(t\) inspects \(u_t\) and observes \(s_t = S(u_t)\), evidence flows from inspected scopes and callers toward candidate repair targets (\(\texttt{parent}\to\texttt{child}\), \(\texttt{caller}\to\texttt{callee}\)) with degree damping \(1/\sqrt{\deg_r(v)}\) (plus \(0.5\)-weighted 2-hop class-entry edges from callers of a class to its methods):
\begin{equation}
  \label{eq:update}
  b_t(v) \;=\; b_{t-1}(v) \;+\; \ell_{\texttt{self}}(s_t)\,\mathbb{I}[v = u_t] \;+\; \sum_{r \in \mathcal{R}_+} \frac{w_r(u_t \to v)}{\sqrt{\max(1, \deg_r(v))}}\,\ell_r(s_t).
\end{equation}
At each step \(t \in \{1,\dots,B\}\), GraphScent queries the unvisited candidate maximizing \(a_t(v) = b_{t-1}(v) + c / \sqrt{1 + n_{\text{vis}}(v)}\), resetting KV cache after every call. On \texttt{E4B}, calibration maps a noisy \(S=3\) caller hit to \(\ell_{\texttt{called\_by}}(3)=+0.26\) while penalizing \(S=0\) at \(-1.39\), preventing false-positive callers from dominating search.

\paragraph{Stage 2: Stateless top-6 listwise commitment.}
Because four-level ordinal scores create ties among plausible candidates, pointwise \(\arg\max\) commitment (\texttt{PW@1}) suffers tie-breaking noise. After \(B\) probes, GraphScent issues one stateless listwise prompt (median \(2.1\text{K}\), P90 \(3.1\text{K}\) input tokens) comparing the top-\(6\) belief-ranked visited candidates side-by-side to select \texttt{LW Commit@1}.

\subsection{Supporting Post-Edit Co-Edit Tool: AST Ripple}
Across \(1{,}087\) single-trajectory \texttt{Gemma 4 31B} issue-resolution runs, solve rates drop \(4.2\times\) from single-file patches (\(39.1\%\), \(n=667\)) to \(2\)-file (\(18.6\%\), \(n=234\)) and \(3+\)-file patches (\(9.4\%\), \(n=186\); Figure~\ref{fig:audit}b). In Stage~3, \textbf{AST Ripple} parses an agent's draft diff, extracts modified function/class signatures, queries \(G\) for unedited direct callers (\texttt{called\_by}) and overrides (\texttt{mirror}) in other files, and formats a compact \(\le 900\)-character (\(\sim 160\)-token) co-edit alert.

\section{Experimental Evaluation}
\label{sec:eval}

\subsection{Experimental Setup, Baselines, and Multi-Hop Stratification}
We evaluate all methods live at \(T=0\) (\texttt{thinking=off}, \texttt{vLLM 0.19.1} on an isolated single NVIDIA H200 NVL GPU; all 4-bit QAT \texttt{w4a16} models fit within \(<21.5\,\text{GiB}\) VRAM) across \texttt{E4B}, \texttt{12B}, and \texttt{31B} on \texttt{public\_113} and \texttt{post2025\_98} (\texttt{pooled\_211}) at budgets \(B=24\) (main text) and \(B=15\) (Appendix~\ref{app:ablations}) using identical folded views and prompts.

\paragraph{Baselines and ablations.}
We compare \texttt{GraphScent} against six baselines—\texttt{Flat BM25} (top-\(B\) lexical candidates + rerank), \texttt{BAGEL-GP (Comp256)} \citep{kim2026bagel} (RBF Gaussian process active search over supplied Oct~13 256-d embeddings on \texttt{public\_113}), \texttt{Oct13-Undir} (score propagation on supplied Oct~13 call graphs), \texttt{Static Dir-PPR} (\texttt{RepoAtlas}-style \citep{zhang2026repoatlas} \texttt{BM25}-personalized directed PageRank top-\(B\) + rerank), \texttt{Undir-AST} (symmetric positive propagation across all Typed-AST edges), and \textbf{\texttt{Fixed-Dir}} (uncalibrated directional search propagating fixed weight \(0.5\,(s/3)\) strictly along \(\texttt{parent}\to\texttt{child}\) and \(\texttt{caller}\to\texttt{callee}\))—plus five single-component ablations (\texttt{Raw-Score}, \texttt{No-Gate}, \texttt{Rev-Dir}, \texttt{No-Damp}, \texttt{No-2Hop}). We track the 3-stage funnel \texttt{Found@24} \(\to\) \texttt{Top6@24} \(\to\) \texttt{PW@1} / \texttt{LW Commit@1} on all \(211\) tasks, on \textbf{\texttt{BM25-Hard}} (best gold \texttt{BM25} rank \(>10\), \(n=94\)), and on \textbf{\texttt{Graph Hop \(\ge 2\)}} (shortest undirected Typed-AST path from \texttt{BM25} top-5 to \(V^*\) is \(\ge 2\) hops, \(n=65\)).

\begin{table}[t]
  \caption{Live \(T=0\) localization funnel at budget \(B=24\) (\texttt{pooled\_211}: \(113\) public + \(98\) post-2025 tasks), plus panel split (\texttt{Pub} / \texttt{Post} \texttt{LW@1}), \texttt{BM25-Hard} (\(\text{rank}>10\), \(n=94\)), and \texttt{Graph Hop \(\ge 2\)} (\(d_{\text{AST}}^{\text{undir}} \ge 2\), \(n=65\)). On \texttt{public\_113} (\texttt{E4B}/\texttt{12B}/\texttt{31B}), \texttt{BAGEL-GP (Comp256)} achieves \(34.5\%/38.1\%/38.9\%\) \texttt{LW@1} and \texttt{Oct13-Undir} achieves \(31.9\%/38.1\%/37.2\%\) vs.\ \texttt{GraphScent} \(39.8\%/37.2\%/45.1\%\). Complete 12-mode tables for \(B \in \{15,24\}\) appear in Appendix~\ref{app:ablations}.}
  \label{tab:main_results}
  \centering
  \small
  \setlength{\tabcolsep}{3.7pt}
  \fitbox{\begin{tabular}{llcccccccc}
    \toprule
    \textbf{Scale} & \textbf{Method (\(B=24\))} & \textbf{\texttt{Found}} & \textbf{\texttt{Top6}} & \textbf{\texttt{PW@1}} & \textbf{\texttt{LW@1 (211)}} & \textbf{Pub / Post} & \textbf{\texttt{BM25>10 (94)}} & \textbf{\texttt{Hop\(\ge\)2 (65)}} \\
    \midrule
    \multirow{7}{*}{\textbf{\texttt{E4B}}}
      & \texttt{Flat BM25}          & \(69.2\%\) & \(57.8\%\) & \(29.4\%\) & \(39.3\%\) & \(39.8 / 38.8\%\) & \(3.2\%\)  & \(12.3\%\) \\
      & \texttt{Static Dir-PPR}     & \(70.1\%\) & \(58.3\%\) & \(28.4\%\) & \(38.9\%\) & \(38.1 / 39.8\%\) & \(5.3\%\)  & \(10.8\%\) \\
      & \texttt{Undir-AST}          & \(\mathbf{70.6\%}\) & \(\mathbf{58.8\%}\) & \(29.4\%\) & \(35.5\%\) & \(33.6 / 37.8\%\) & \(4.3\%\)  & \(9.2\%\)  \\
      & \texttt{Fixed-Dir} (uncal.) & \(66.8\%\) & \(57.8\%\) & \(\mathbf{30.3\%}\) & \(37.4\%\) & \(36.3 / 38.8\%\) & \(6.4\%\)  & \(6.2\%\)  \\
      & \texttt{Ablate: Raw-Score}  & \(66.8\%\) & \(54.5\%\) & \(23.7\%\) & \(37.4\%\) & \(31.0 / \mathbf{44.9\%}\) & \(\mathbf{13.8\%}\) & \(\mathbf{13.8\%}\) \\
      & \texttt{Ablate: Rev-Dir}    & \(64.0\%\) & \(57.3\%\) & \(29.9\%\) & \(37.0\%\) & \(34.5 / 39.8\%\) & \(3.2\%\)  & \(9.2\%\)  \\
      & \textbf{\texttt{GraphScent}} & \(\mathbf{70.6\%}\) & \(\mathbf{58.8\%}\) & \(28.9\%\) & \(\mathbf{41.2\%}\) & \(\mathbf{39.8} / 42.9\%\) & \(10.6\%\) & \(\mathbf{13.8\%}\) \\
    \midrule
    \multirow{7}{*}{\textbf{\texttt{12B}}}
      & \texttt{Flat BM25}          & \(69.2\%\) & \(63.0\%\) & \(\mathbf{35.5\%}\) & \(39.3\%\) & \(37.2 / 41.8\%\) & \(9.6\%\)  & \(16.9\%\) \\
      & \texttt{Static Dir-PPR}     & \(70.1\%\) & \(64.0\%\) & \(32.7\%\) & \(39.8\%\) & \(37.2 / 42.9\%\) & \(10.6\%\) & \(16.9\%\) \\
      & \texttt{Undir-AST}          & \(71.6\%\) & \(65.4\%\) & \(35.1\%\) & \(41.7\%\) & \(39.8 / 43.9\%\) & \(17.0\%\) & \(\mathbf{20.0\%}\) \\
      & \texttt{Fixed-Dir} (uncal.) & \(70.1\%\) & \(64.9\%\) & \(35.1\%\) & \(\mathbf{44.1\%}\) & \(\mathbf{43.4} / 44.9\%\) & \(\mathbf{20.2\%}\) & \(\mathbf{20.0\%}\) \\
      & \texttt{Ablate: Raw-Score}  & \(\mathbf{72.0\%}\) & \(\mathbf{65.9\%}\) & \(30.8\%\) & \(41.7\%\) & \(37.2 / \mathbf{46.9\%}\) & \(16.0\%\) & \(15.4\%\) \\
      & \texttt{Ablate: Rev-Dir}    & \(69.2\%\) & \(63.5\%\) & \(\mathbf{35.5\%}\) & \(39.3\%\) & \(38.1 / 40.8\%\) & \(7.4\%\)  & \(15.4\%\) \\
      & \textbf{\texttt{GraphScent}} & \(71.1\%\) & \(65.4\%\) & \(29.9\%\) & \(41.7\%\) & \(37.2 / \mathbf{46.9\%}\) & \(17.0\%\) & \(15.4\%\) \\
    \midrule
    \multirow{7}{*}{\textbf{\texttt{31B}}}
      & \texttt{Flat BM25}          & \(69.2\%\) & \(64.5\%\) & \(36.0\%\) & \(41.7\%\) & \(38.9 / 44.9\%\) & \(12.8\%\) & \(18.5\%\) \\
      & \texttt{Static Dir-PPR}     & \(70.1\%\) & \(64.9\%\) & \(35.1\%\) & \(42.2\%\) & \(38.9 / \mathbf{45.9\%}\) & \(13.8\%\) & \(18.5\%\) \\
      & \texttt{Undir-AST}          & \(\mathbf{74.9\%}\) & \(68.2\%\) & \(36.0\%\) & \(41.7\%\) & \(39.8 / 43.9\%\) & \(17.0\%\) & \(15.4\%\) \\
      & \texttt{Fixed-Dir} (uncal.) & \(74.4\%\) & \(67.8\%\) & \(35.5\%\) & \(43.6\%\) & \(42.5 / 44.9\%\) & \(20.2\%\) & \(20.0\%\) \\
      & \texttt{Ablate: No-Gate}    & \(\mathbf{75.4\%}\) & \(\mathbf{69.2\%}\) & \(36.0\%\) & \(\mathbf{44.5\%}\) & \(\mathbf{46.9} / 41.8\%\) & \(\mathbf{24.5\%}\) & \(\mathbf{23.1\%}\) \\
      & \texttt{Ablate: Rev-Dir}    & \(72.0\%\) & \(66.4\%\) & \(\mathbf{37.0\%}\) & \(43.1\%\) & \(40.7 / \mathbf{45.9\%}\) & \(14.9\%\) & \(20.0\%\) \\
      & \textbf{\texttt{GraphScent}} & \(74.9\%\) & \(68.7\%\) & \(36.0\%\) & \(43.6\%\) & \(45.1 / 41.8\%\) & \(23.4\%\) & \(21.5\%\) \\
    \bottomrule
  \end{tabular}}
\end{table}

\begin{figure}[t]
  \centering
  \includegraphics[width=\textwidth]{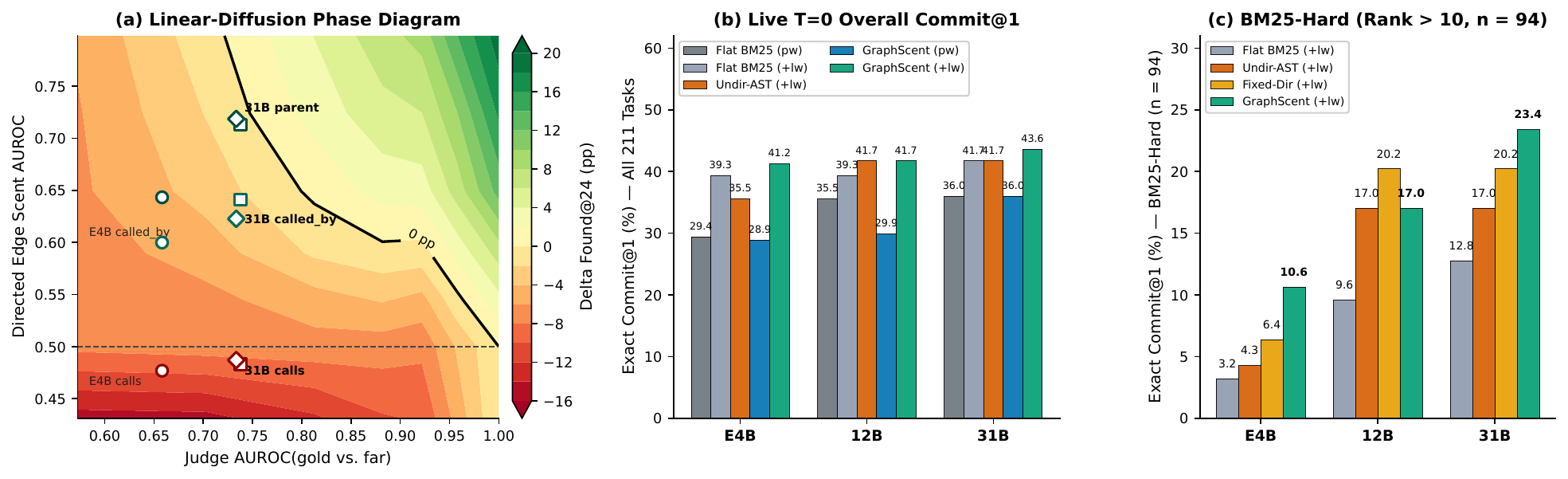}
  \caption{\textbf{Phase diagram and live \(T=0\) localization results.} \textbf{(a)} \(5{,}760\)-run linear-diffusion phase diagram showing \(\Delta\texttt{Found@24}\) (percentage points) across direct judge \(\mathrm{AUROC}(\texttt{gold}\;\text{vs.}\;\texttt{far})\) and neighbor edge scent, overlaid with empirical Gemma~4 operating points. \textbf{(b)} Overall \texttt{Commit@1} across all \(211\) tasks (\(B=24\)). \textbf{(c)} \texttt{BM25-Hard} (\(\text{rank}>10\), \(n=94\)) \texttt{Commit@1} comparing \texttt{Flat BM25}, \texttt{Undir-AST}, \texttt{Fixed-Dir}, and \texttt{GraphScent}.}
  \label{fig:results}
\end{figure}

\subsection{Main Localization Findings and Mechanism Decomposition}
Table~\ref{tab:main_results} and Figure~\ref{fig:results} clarify what each component contributes across scales.

\paragraph{Forward orientation drives non-lexical target recovery.}
Reversing propagation direction (\texttt{Ablate: Rev-Dir}: \(\texttt{callee}\to\texttt{caller}\), \(\texttt{child}\to\texttt{parent}\)) lowers \texttt{Found@24} at every scale (\(70.6\% \to 64.0\%\) on \texttt{E4B}, \(71.1\% \to 69.2\%\) on \texttt{12B}, \(74.9\% \to 72.0\%\) on \texttt{31B}) and sharply reduces \texttt{BM25-Hard} \texttt{LW@1} (\(10.6\% \to 3.2\%\) on \texttt{E4B}, \(17.0\% \to 7.4\%\) on \texttt{12B}, \(23.4\% \to 14.9\%\) on \texttt{31B}). At \texttt{31B}, \texttt{GraphScent} improves \texttt{BM25-Hard} \texttt{LW@1} by \(+10.6\,\text{pp}\) over \texttt{Flat BM25} (\(23.4\%\) vs.\ \(12.8\%\), paired bootstrap \(95\%\) CI \([+2.1, +19.1\,\text{pp}]\), \(p=0.015\), \(14\) wins vs.\ \(4\) losses) and improves \texttt{Graph Hop \(\ge 2\)} \texttt{LW@1} to \(21.5\%\) vs.\ \(18.5\%\) (\texttt{Flat BM25}) and \(15.4\%\) (\texttt{Undir-AST}; reaching \(26.2\%\) under 1-hop \texttt{Ablate: No-2Hop}). Across all \(211\) tasks—where \(55.5\%\) already contain a gold node inside \texttt{BM25} top-10—overall gains over \texttt{Flat BM25} are naturally more modest (\(+1.9\) to \(+2.4\,\text{pp}\)).

\paragraph{When out-of-repository \texttt{LLR} calibration matters over \texttt{Fixed-Dir}.}
Comparing \texttt{GraphScent} against uncalibrated \texttt{Fixed-Dir} reveals a clear scale interaction. On \textbf{\texttt{E4B}}, where \(50.9\%\) of \texttt{public\_113} distractors receive \(S=3\), uncalibrated \texttt{Fixed-Dir} (\(66.8\%\) \texttt{Found@24}, \(37.4\%\) \texttt{LW@1}) and \texttt{Ablate: Raw-Score} (\(66.8\%\) \texttt{Found@24}, \(37.4\%\) \texttt{LW@1}) chase false-positive callers; out-of-repository \texttt{LLR} calibration lifts \texttt{E4B} to \(70.6\%\) \texttt{Found@24} and \(41.2\%\) \texttt{LW@1} (\(+3.8\,\text{pp}\) over \texttt{Fixed-Dir}; \(+5.7\,\text{pp}\) over \texttt{Undir-AST}, \(95\%\) CI \([+0.5, +10.4\,\text{pp}]\), \(p=0.037\); and \(+7.4\,\text{pp}\) over \texttt{Flat BM25} on \texttt{BM25-Hard}, \(p=0.033\)). Conversely, at \textbf{\texttt{12B}} and \textbf{\texttt{31B}}, where distractor \(S=3\) rates fall below \(18\%\) and \(10\%\), simple \texttt{Fixed-Dir} already captures most overall directional gains (\(44.1\%\) on \texttt{12B}, \(43.6\%\) on \texttt{31B}), while \texttt{GraphScent} retains a \(+3.2\,\text{pp}\) advantage on \texttt{31B} \texttt{BM25-Hard} (\(23.4\%\) vs.\ \(20.2\%\)). Similarly, \texttt{Ablate: No-Gate} (\(44.5\%\) on \texttt{31B}) indicates that once \texttt{31B}'s fitted \texttt{LLR}s automatically assign near-zero weight to \texttt{calls} (\(\ell_{\texttt{calls}}(3)=-0.07\)), soft \texttt{LLR} weighting can exploit occasional sub-threshold \texttt{mirror} edges without hard \(\tau=0.55\) pruning.

\paragraph{Candidate-pool hygiene, listwise tie-breaking, and token footprint.}
Tracking \texttt{Found@24 \(\to\) Top6@24 \(\to\) LW@1} explains why symmetric \texttt{Undir-AST} stalls at commitment despite matching discovery (\texttt{Found@24} \(= 70.6\%\) on \texttt{E4B}, \(74.9\%\) on \texttt{31B}): boosting \(2.4\times\) downstream callees inflates non-target posterior scores, dropping \texttt{Undir-AST} \texttt{LW@1} to \(35.5\%\) (\texttt{E4B}) and \(41.7\%\) (\texttt{31B}), and \(4.3\%\) and \(17.0\%\) on \texttt{BM25-Hard}. Over GraphScent's cleaner \texttt{Top6@24} shortlist (\(40.4\%\) vs.\ \(24.5\%\) \texttt{Flat BM25} on \texttt{31B} \texttt{BM25-Hard}), the single stateless listwise call improves \texttt{LW@1} over pointwise \texttt{PW@1} by \(+12.3\,\text{pp}\) (\texttt{E4B}, \(p<0.001\)), \(+11.8\,\text{pp}\) (\texttt{12B}, \(p<0.001\)), and \(+7.6\,\text{pp}\) (\texttt{31B}, \(p=0.004\)). Stateless node-scoring prompts average a median of \(657\text{--}691\) input tokens (P90 \(1{,}266\text{--}1{,}323\)) and the top-6 listwise prompt uses a median of \(2{,}050\text{--}2{,}197\) tokens (P90 \(3{,}086\text{--}3{,}147\)), totaling a median of \(15.2\text{K}\text{--}18.9\text{K}\) cumulative input tokens (\(\sim 225\) output tokens) across \(25\) calls per task while keeping single-GPU VRAM at \(10.42\), \(8.51\), and \(20.28\,\text{GiB}\).

\subsection{Supporting Post-Edit Evaluation: AST Ripple Reference-File Recall}
We evaluate \texttt{AST Ripple} retrospectively across three single-trajectory agent runs: two independent non-zero-temperature \texttt{Gemma 4 31B} runs on \texttt{post2025\_102} (\texttt{Run A}: \(35/72 = 48.6\%\) single-file vs.\ \(4/30 = 13.3\%\) multi-file resolved; \texttt{Run B}: \(31/72 = 43.1\%\) single-file vs.\ \(3/30 = 10.0\%\) multi-file) and a \(248\)-task \texttt{Gemma 4 E4B} panel (\(35/162 = 21.6\%\) single-file vs.\ \(9/86 = 10.5\%\) multi-file). On unresolved multi-file tasks where the agent's patch missed at least one reference-patch file, \texttt{AST Ripple} (median \(618\text{--}678\) chars on \texttt{31B}, \(365\) on \texttt{E4B}) surfaces a missed reference file in \(\mathbf{50.0\%}\) (\(11/22\)) of tasks in both \texttt{31B} runs and \(\mathbf{29.6\%}\) (\(21/71\)) on \texttt{E4B}. A zero-shot Gemma~4 ripple judge (\(33\) missed reference targets vs.\ \(82\) unedited neighbors across \(31\) tasks) ranks missed reference files at \(\mathrm{AUROC}\) \(0.647\;[.56,.74]\) (\texttt{E4B}), \(0.659\;[.53,.77]\) (\texttt{12B}), and \(0.582\;[.51,.65]\) (\texttt{31B}). These measurements establish compact missed-reference-file recall rather than closed-loop test-suite improvement, which remains future work.

\section{Related Work}
\label{sec:related}

\paragraph{Graph-guided code localization and search agents.}
LocAgent \citep{chen2025locagent}, CoSIL \citep{jiang2025cosil}, and OrcaLoca \citep{yu2025orcaloca} construct directed heterogeneous repository graphs (\texttt{contain}, \texttt{invoke}, \texttt{inherit}) navigated by multi-turn LLM agents, while LogicLoc \citep{xu2026logicloc}, LARGER \citep{hu2026larger}, SweRank \citep{reddy2025swerank}, GREPO \citep{wang2026grepo}, and SWE-Explore \citep{zhang2026sweexplore} investigate neurosymbolic localization, lexical-graph retrieval, code reranking, GNN benchmarks, and exploration dynamics. Most closely related on the static-graph side, RepoAtlas \citep{zhang2026repoatlas} builds directed call and inheritance views ranked via Personalized PageRank. On the policy-learning side, \(\text{A}^2\text{Agent}\) \citep{kim2026a2agent}, RepoNavigator \citep{zhang2025reponavigator}, CodeScout \citep{sutawika2026codescout}, and ToolTrain \citep{ma2025tooltrain} train multi-step search policies via reinforcement learning on corpora such as SWE-Gym \citep{pan2025swegym} and SWE-smith \citep{yang2025swesmith}, complementing workflow scaffolds like Agentless \citep{xia2024agentless}. Rather than proposing directed code graphs or policy training, our work isolates which structural directions make unedited LLM relevance judgments diagnostic (\texttt{parent} and \texttt{called\_by} vs.\ \texttt{calls}), evaluates \texttt{Static Dir-PPR} and \texttt{Fixed-Dir} controls directly, and demonstrates how out-of-repository calibration stabilizes small-model bounded-context search.

\paragraph{Bayesian active retrieval and graph belief propagation.}
Classical Bayesian optimal active search \citep{garnett2012bayesian} and information foraging theory \citep{pirolli1999information} formalize sequential target discovery from proximal cues. Recently, BAGEL \citep{kim2026bagel} and ReBOL \citep{korikov2026rebol} guide dense passage retrieval by fitting query-specific Gaussian processes to sparse LLM relevance scores, while ALPER \citep{wang2026alper} and EoG \citep{jha2026eog} pair local LLM judgments with graph label or belief propagation for entity resolution and root-cause analysis. GraphScent adapts active observation to typed AST multigraphs: rather than smoothing isotropically in embedding space—where supplied 256-d competition neighbors score \(\mathrm{Scent}=0.436\text{--}0.470\)—it propagates empirically calibrated log-likelihood ratios selectively along positive-scent directed relations.

\section{Discussion, Limitations, and Reproducibility}
\label{sec:discussion}

\paragraph{Scope and limitations.}
Our study has four explicit boundaries. First, all experiments cover Python repositories (\(211\) tasks across \(86\) repos, excluding \(20\) tasks whose reference patches modify only module-level code, new symbols, or non-Python files); statically typed languages with explicit type signatures may exhibit different caller-versus-callee asymmetries. Second, we evaluate three parameter scales (\texttt{E4B}, \texttt{12B}, \texttt{31B}) of a single model family (Gemma~4) under \(\le 60\)-line folded views and a fixed \(T=0\) rubric. Third, while \texttt{post2025\_98} (\(2025\text{--}2026\)) has zero repository overlap with \texttt{public\_113}, timestamps alone cannot guarantee pretraining non-exposure. Fourth, \texttt{AST Ripple} is evaluated on missed reference-patch file recall rather than end-to-end test-suite resolution. Appendix~\ref{app:negative} also documents negative results, including rank-16 commit-patch LoRA fine-tuning—which degraded 1-hop neighbor scent (\(\mathrm{AUROC}\;0.570 \to 0.536\)) by teaching binary rejection of unedited callers—and \texttt{thinking=on}, which lowered JSON parse reliability (\(100.0\% \to 96.0\%\)) under bounded output budgets without improving neighbor scent.

\paragraph{Reproducibility and open resource.}
All artifacts are publicly released under Apache-2.0:
\begin{itemize}[leftmargin=1.6em,itemsep=1.5pt,topsep=2pt]
  \item \textbf{Dataset \& Toolkit (\texttt{GraphScent-Bench}):} \url{https://www.kaggle.com/datasets/rameshln/graphscent-bench-gemma4} (standalone \texttt{graphscent} Python package, audited manifests for all \(211\) tasks, \(46{,}032\) cached live \(T=0\) node judgments and listwise commits across \texttt{E4B}/\texttt{12B}/\texttt{31B}, out-of-repository calibration folds, and unit tests).
  \item \textbf{Executable Reproduction Notebook:} \url{https://www.kaggle.com/code/rameshln/graphscent-bench-repro} (regenerates every table, confidence interval, and figure in under \(10\,\text{s}\) on CPU).
  \item \textbf{Benchmark Track Overview:} \url{https://www.kaggle.com/competitions/gemma-4-developer-agent-paper}.
\end{itemize}

\paragraph{AI-assistance disclosure.}
In accordance with the Gemma~4 Paper Track rules, AI coding and writing assistants were used to help prototype evaluation scripts, run automated statistical verification checks, format figures and \LaTeX{} tables, and polish prose clarity. All experimental designs, hypotheses, live model evaluations, control experiments, and scientific claims were conceived, inspected, and verified against raw experimental logs by the author.

\bibliographystyle{plainnat}
\bibliography{refs}

\newpage
\appendix

\section{Algorithmic Details and Folded View Construction}
\label{app:algorithm}

\paragraph{Runtime Typed-AST multigraph construction.}
For a repository snapshot, \texttt{graphscent.ast\_graph} walks all non-test \texttt{.py} files (excluding \texttt{tests/}, \texttt{test\_*.py}, \texttt{docs/}, and virtual-environment directories) via Python's standard \texttt{ast} module. Every module, class, function, and async function becomes a node \(v \in V\) keyed by \texttt{relpath::QualName}. Four directed edge families are extracted in \(O(|V| + |E|)\) time: (1) lexical containment (\texttt{parent}/\texttt{child} and induced peer \texttt{sibling} relations within a class); (2) resolved intra-repository calls (\texttt{calls}/\texttt{called\_by}), resolving \texttt{self.m()} and \texttt{cls.m()} within \(u\)'s class MRO, imported module/symbol aliases, and unambiguous repository-wide symbol names; (3) class inheritance (\texttt{inherits}) between in-repository base and derived classes; and (4) parallel method overrides (\texttt{mirror}) linking non-dunder methods sharing an identical name across classes related by inheritance or shared protocol interfaces.

\paragraph{Deterministic \(\le 60\)-line folded code views.}
To bound per-call prompt tokens regardless of entity length, each node \(u\) is rendered into at most \(60\) lines: functions/methods \(\le 60\) lines are rendered verbatim with 1-based line numbers; functions/methods \(> 60\) lines retain decorators, signature, docstring, control-flow headers (\texttt{if}/\texttt{for}/\texttt{while}/\texttt{try}/\texttt{except}), call statements, and \texttt{return}/\texttt{raise} lines up to \(60\) lines; and classes \(> 60\) lines (\(n=114\) of \(187\) sampled \texttt{parent} nodes) are rendered as class skeletons preserving the class signature, class docstring, class-level attributes, and 1-line \texttt{def} headers of child methods.

\paragraph{Active search update rules.}
Given budget \(B=24\), lexical scaling \(\alpha=3.0\) on min-max normalized \texttt{BM25} scores, exploration constant \(c=0.35\), gate \(\tau=0.55\), and out-of-repository log-likelihood ratio table \(\{\ell_r(s)\}\):
\begin{enumerate}[leftmargin=1.6em,itemsep=2pt,topsep=2pt]
  \item Initialize \(b_0(v) = \alpha\,\widetilde{\mathrm{BM25}}(q,v)\) for all candidate functions and classes \(v \in V\), visited set \(\mathcal{V}_0 = \emptyset\), and neighbor-visit count \(n_{\text{vis}}(v) = 0\).
  \item For \(t = 1, \dots, B\): select \(u_t = \arg\max_{v \in V \setminus \mathcal{V}_{t-1}} [\,b_{t-1}(v) + c / \sqrt{1 + n_{\text{vis}}(v)}\,]\) (guarding the top-3 raw \texttt{BM25} seeds within the first three probes); query the frozen local LLM at \(T=0\) (\texttt{thinking=off}) on the folded view of \(u_t\) to obtain \(s_t = S(u_t) \in \{0,1,2,3\}\); add self-evidence \(\ell_{\texttt{self}}(s_t)\) to \(b_t(u_t)\); and for each gated relation \(r \in \mathcal{R}_+\) and target candidate \(v\) standing in diagnostic relation \(u_t \xrightarrow{r} v\) (\(\texttt{parent}\to\texttt{child}\), \(\texttt{caller}\to\texttt{callee}\), \(\texttt{inherits}\), \(\texttt{mirror}\), plus \(0.5\)-weighted 2-hop class-entry edges from callers of class \(c\) to methods \(m \in \texttt{child}(c)\)), increment \(b_t(v)\) by \(w_r(u_t \to v)\,\ell_r(s_t)/\sqrt{\max(1,\deg_r(v))}\) and increment \(n_{\text{vis}}(v)\).
  \item Rank visited nodes \(v \in \mathcal{V}_B\) by \((b_B(v), S(v), \widetilde{\mathrm{BM25}}(q,v))\); pass the top-\(6\) folded candidates to one stateless listwise \(T=0\) prompt to output \(\hat{v}_{\text{LW}}\) (\texttt{LW Commit@1}).
\end{enumerate}

\FloatBarrier
\section{Out-of-Repository LLR Calibration and Threshold Sensitivity}
\label{app:calibration}

Table~\ref{tab:llrs} lists the exact Laplace-smoothed out-of-repository log-likelihood ratios \(\ell_r(s) = \log[\widehat{P}(S=s \mid r)/\widehat{P}(S=s \mid \texttt{far})]\) fitted on \texttt{public\_113} (transferred zero-shot without modification to all \(82\) repositories in \texttt{post2025\_98}) alongside raw distractor score histograms. On \texttt{E4B}, because \(50.9\%\) of \texttt{far} distractors receive \(S=3\), an \(S=3\) score on a caller receives a tempered update (\(+0.26\)), whereas \(S=2\) (\(+0.69\)) and the absence of \(S=0\) (\(-1.39\)) carry sharp discriminative leverage. On \texttt{12B} and \texttt{31B}, where \texttt{far} distractors receive \(S=0\) at \(64.0\%\) and \(82.7\%\), intermediate execution-path scores \(S=1\) and \(S=2\) on callers carry strong positive log-likelihood ratios (\(+0.70\) to \(+1.35\)), while \texttt{calls} receives near-zero weight (\(-0.04\) and \(-0.07\) at \(S=3\)). Across leave-one-repository-out folds on \texttt{public\_113} and across gate thresholds \(\tau \in \{0.52, 0.55, 0.58\}\), \texttt{parent} and \texttt{called\_by} are selected into \(\mathcal{R}_+\) in \(14/15\) scale-threshold combinations (\(100\%\) at default \(\tau=0.55\)), whereas \texttt{calls} and \texttt{knear\_comp256} are excluded in \(15/15\) combinations.

\begin{table}[!htbp]
  \caption{Fitted Laplace-smoothed log-likelihood ratios \(\ell_r(s)\) for \(s \in \{0,1,2,3\}\) on \texttt{public\_113} (used zero-shot on \texttt{post2025\_98}) and empirical \texttt{far}-distractor score distributions across Gemma~4 scales.}
  \label{tab:llrs}
  \centering
  \small
  \begin{tabular}{llrrrrr}
    \toprule
    \textbf{Scale} & \textbf{Relation / Distribution} & \(\mathrm{Scent}_{\text{pub}}\) & \(s = 0\) & \(s = 1\) & \(s = 2\) & \(s = 3\) \\
    \midrule
    \multirow{5}{*}{\textbf{\texttt{E4B}}}
      & \(\widehat{P}(S=s \mid \texttt{far})\) (\(n=678\)) & \(0.500\) & \(28.2\%\) & \(17.6\%\) & \(3.4\%\) & \(50.9\%\) \\
      & \texttt{parent} \(\ell_{\texttt{parent}}(s)\)      & \(0.598\) & \(-0.46\) & \(-0.77\) & \(+0.39\) & \(+0.31\) \\
      & \texttt{called\_by} \(\ell_{\texttt{called\_by}}(s)\) & \(0.613\) & \(-1.39\) & \(+0.13\) & \(+0.69\) & \(+0.26\) \\
      & \texttt{inherits} \(\ell_{\texttt{inherits}}(s)\)  & \(0.600\) & \(-0.34\) & \(-0.57\) & \(+1.74\) & \(-0.01\) \\
      & \texttt{calls} \(\ell_{\texttt{calls}}(s)\) (excluded) & \(0.462\) & \(-0.04\) & \(+0.49\) & \(+0.25\) & \(-0.24\) \\
    \midrule
    \multirow{5}{*}{\textbf{\texttt{12B}}}
      & \(\widehat{P}(S=s \mid \texttt{far})\) (\(n=678\)) & \(0.500\) & \(64.0\%\) & \(15.5\%\) & \(3.1\%\) & \(17.4\%\) \\
      & \texttt{parent} \(\ell_{\texttt{parent}}(s)\)      & \(0.684\) & \(-0.65\) & \(-0.27\) & \(+0.83\) & \(+1.00\) \\
      & \texttt{called\_by} \(\ell_{\texttt{called\_by}}(s)\) & \(0.683\) & \(-0.86\) & \(+0.70\) & \(+0.92\) & \(+0.65\) \\
      & \texttt{mirror} \(\ell_{\texttt{mirror}}(s)\)      & \(0.582\) & \(-0.37\) & \(+0.55\) & \(+0.82\) & \(+0.23\) \\
      & \texttt{calls} \(\ell_{\texttt{calls}}(s)\) (excluded) & \(0.478\) & \(+0.07\) & \(-0.34\) & \(+0.13\) & \(-0.04\) \\
    \midrule
    \multirow{5}{*}{\textbf{\texttt{31B}}}
      & \(\widehat{P}(S=s \mid \texttt{far})\) (\(n=678\)) & \(0.500\) & \(82.7\%\) & \(6.5\%\)  & \(1.2\%\) & \(9.6\%\) \\
      & \texttt{parent} \(\ell_{\texttt{parent}}(s)\)      & \(0.696\) & \(-0.58\) & \(-0.80\) & \(+1.50\) & \(+1.54\) \\
      & \texttt{called\_by} \(\ell_{\texttt{called\_by}}(s)\) & \(0.651\) & \(-0.48\) & \(+1.35\) & \(+0.72\) & \(+0.78\) \\
      & \texttt{inherits} \(\ell_{\texttt{inherits}}(s)\)  & \(0.652\) & \(-0.72\) & \(+1.51\) & \(+2.03\) & \(+0.73\) \\
      & \texttt{calls} \(\ell_{\texttt{calls}}(s)\) (excluded) & \(0.478\) & \(+0.05\) & \(-0.70\) & \(-0.07\) & \(-0.07\) \\
    \bottomrule
  \end{tabular}
\end{table}

\FloatBarrier
\section{Detailed Confound, Masked-Parent, and Lexical-Stratum Controls}
\label{app:confounds}

Table~\ref{tab:confounds_full} breaks down the zero-leakage and lexical-overlap controls across \texttt{public\_113} and \texttt{post2025\_98} and across three \texttt{BM25} lexical-similarity strata (\(\widetilde{\mathrm{BM25}} < 0.20\), \([0.20, 0.45)\), and \(\ge 0.45\)). Across every panel and every \texttt{BM25} band, \texttt{called\_by} outranks \texttt{calls} (\(\mathrm{AUROC} > 0.55\) in all nine scale-by-stratum cells) and outranks \texttt{far} distractors within matched \texttt{BM25} bands (\(\mathrm{AUROC}\;0.650\text{--}0.756\)). Because \texttt{far} distractors are drawn from top-\texttt{BM25} candidates (mean \(\widetilde{\mathrm{BM25}} = 0.634\) vs.\ \(0.308\) for \texttt{called\_by}), the unstratified \(\mathrm{Scent}(\texttt{called\_by}) = 0.600\text{--}0.641\) in Table~\ref{tab:scent} is a conservative lower bound.

\begin{table}[!htbp]
  \caption{Panel-by-panel zero-leakage masked-\texttt{parent} controls and \texttt{BM25}-stratified span-disjoint \texttt{called\_by} controls across \texttt{E4B}, \texttt{12B}, and \texttt{31B}.}
  \label{tab:confounds_full}
  \centering
  \small
  \setlength{\tabcolsep}{4.5pt}
  \fitbox{\begin{tabular}{llccc}
    \toprule
    \textbf{Control Subset / Stratum} & \textbf{Counts} & \textbf{\texttt{E4B}} & \textbf{\texttt{12B}} & \textbf{\texttt{31B}} \\
    \midrule
    \texttt{public\_113}: Original \texttt{parent} vs.\ \texttt{far}        & \(131\) vs.\ \(678\) & \(0.598\;[.55,.65]\) & \(0.684\;[.63,.75]\) & \(0.696\;[.64,.75]\) \\
    \texttt{public\_113}: \texttt{parent\_gold\_masked} vs.\ \texttt{far}   & \(131\) vs.\ \(678\) & \(0.612\;[.55,.66]\) & \(0.673\;[.62,.73]\) & \(0.675\;[.62,.73]\) \\
    \texttt{public\_113}: \texttt{parent\_header\_doc} vs.\ \texttt{far}    & \(131\) vs.\ \(678\) & \(0.613\;[.56,.67]\) & \(0.685\;[.62,.75]\) & \(0.714\;[.66,.77]\) \\
    \midrule
    \texttt{post2025\_98}: Original \texttt{parent} vs.\ \texttt{far}       & \(56\) vs.\ \(588\)  & \(0.688\;[.61,.77]\) & \(0.768\;[.69,.85]\) & \(0.775\;[.70,.85]\) \\
    \texttt{post2025\_98}: \texttt{parent\_gold\_masked} vs.\ \texttt{far}  & \(56\) vs.\ \(588\)  & \(0.688\;[.60,.78]\) & \(0.753\;[.67,.84]\) & \(0.764\;[.69,.84]\) \\
    \texttt{post2025\_98}: \texttt{parent\_header\_doc} vs.\ \texttt{far}   & \(56\) vs.\ \(588\)  & \(0.689\;[.61,.78]\) & \(0.776\;[.70,.86]\) & \(0.764\;[.69,.84]\) \\
    \midrule
    Low \texttt{BM25} (\(<0.20\)): \texttt{called\_by} vs.\ \texttt{calls}  & \(119\) vs.\ \(262\) & \(0.633\;[.57,.69]\) & \(0.625\;[.57,.68]\) & \(0.597\;[.55,.64]\) \\
    Mid \texttt{BM25} (\([0.20,0.45)\)): \texttt{called\_by} vs.\ \texttt{calls} & \(120\) vs.\ \(70\) & \(0.606\;[.53,.68]\) & \(0.599\;[.52,.68]\) & \(0.569\;[.49,.65]\) \\
    High \texttt{BM25} (\(\ge 0.45\)): \texttt{called\_by} vs.\ \texttt{calls} & \(58\) vs.\ \(40\) & \(0.554\;[.45,.65]\) & \(0.566\;[.47,.66]\) & \(0.563\;[.44,.67]\) \\
    Mid \texttt{BM25} (\([0.20,0.45)\)): \texttt{called\_by} vs.\ \texttt{far} & \(120\) vs.\ \(291\) & \(0.679\;[.62,.74]\) & \(0.722\;[.66,.78]\) & \(0.703\;[.64,.76]\) \\
    High \texttt{BM25} (\(\ge 0.45\)): \texttt{called\_by} vs.\ \texttt{far} & \(58\) vs.\ \(964\)  & \(0.650\;[.60,.70]\) & \(0.756\;[.70,.81]\) & \(0.756\;[.69,.82]\) \\
    \midrule
    OLS \(\hat{\beta}_{\text{BM25}}\) / \(\hat{\beta}_{\text{len60}}\) (\(n=669\)) & --- & \(+1.07 / -0.13\) & \(+1.57 / +0.10\) & \(+1.53 / +0.03\) \\
    \bottomrule
  \end{tabular}}
\end{table}

\FloatBarrier
\section{Complete 12-Method Localization Funnel, Budgets, and Per-Repository Breakdown}
\label{app:ablations}

Table~\ref{tab:full_ablations} reports all \(10\) pooled methods (\(12\) on \texttt{public\_113} including \texttt{BAGEL-GP} and \texttt{Oct13-Undir}) across \texttt{E4B}, \texttt{12B}, and \texttt{31B} at both \(B=24\) and \(B=15\), including loose file/enclosing-scope commitment (\texttt{LW-Loose}), undirected graph multi-hop (\(d_{\text{AST}}^{\text{undir}} \ge 2\), \(n=65\)), and directed positive-scent multi-hop (\(d_{\text{AST}}^{\text{dir}} \ge 2\), \(n=72\)). Table~\ref{tab:paired_cis} reports paired bootstrap differences, \(95\%\) CIs, and task-level \texttt{Win/Loss/Tie} counts, and Table~\ref{tab:per_repo} breaks down \texttt{public\_113} by repository at natural column widths.

\begin{table}[!htbp]
  \caption{Complete live \(T=0\) localization results across all methods and single-component ablations on \texttt{pooled\_211} at \(B=24\) and \(B=15\).}
  \label{tab:full_ablations}
  \centering
  \footnotesize
  \setlength{\tabcolsep}{3.3pt}
  \fitbox{\begin{tabular}{llccccccccc}
    \toprule
    \textbf{Scale} & \textbf{Method} & \texttt{F@24} & \texttt{T6@24} & \texttt{PW@24} & \texttt{LW@24} & \texttt{LW-Loose} & \texttt{F@15} & \texttt{LW@15} & \texttt{BM25>10} & \texttt{HopU / HopD} \\
    \midrule
    \multirow{10}{*}{\textbf{\texttt{E4B}}}
      & \texttt{Flat BM25}          & \(69.2\%\) & \(57.8\%\) & \(29.4\%\) & \(39.3\%\) & \(46.9\%\) & \(63.5\%\) & \(38.4\%\) & \(3.2\%\)  & \(12.3 / 9.7\%\) \\
      & \texttt{Static Dir-PPR}     & \(70.1\%\) & \(58.3\%\) & \(28.4\%\) & \(38.9\%\) & \(45.5\%\) & \(62.6\%\) & \(37.4\%\) & \(5.3\%\)  & \(10.8 / 8.3\%\) \\
      & \texttt{Undir-AST}          & \(70.6\%\) & \(58.8\%\) & \(29.4\%\) & \(35.5\%\) & \(41.7\%\) & \(61.1\%\) & \(34.6\%\) & \(4.3\%\)  & \(9.2 / 8.3\%\) \\
      & \texttt{Fixed-Dir}          & \(66.8\%\) & \(57.8\%\) & \(30.3\%\) & \(37.4\%\) & \(42.2\%\) & \(64.0\%\) & \(38.4\%\) & \(6.4\%\)  & \(6.2 / 5.6\%\) \\
      & \texttt{Ablate: Raw-Score}  & \(66.8\%\) & \(54.5\%\) & \(23.7\%\) & \(37.4\%\) & \(39.8\%\) & \(62.6\%\) & \(40.8\%\) & \(13.8\%\) & \(13.8 / 12.5\%\) \\
      & \texttt{Ablate: No-Gate}    & \(70.6\%\) & \(58.3\%\) & \(25.6\%\) & \(40.8\%\) & \(43.6\%\) & \(62.6\%\) & \(38.9\%\) & \(10.6\%\) & \(15.4 / 12.5\%\) \\
      & \texttt{Ablate: Rev-Dir}    & \(64.0\%\) & \(57.3\%\) & \(29.9\%\) & \(37.0\%\) & \(44.1\%\) & \(58.8\%\) & \(35.1\%\) & \(3.2\%\)  & \(9.2 / 6.9\%\) \\
      & \texttt{Ablate: No-Damp}    & \(71.1\%\) & \(59.2\%\) & \(28.0\%\) & \(42.2\%\) & \(44.6\%\) & \(64.5\%\) & \(38.9\%\) & \(13.8\%\) & \(13.8 / 11.1\%\) \\
      & \texttt{Ablate: No-2Hop}    & \(70.6\%\) & \(58.3\%\) & \(28.0\%\) & \(40.8\%\) & \(44.1\%\) & \(64.9\%\) & \(39.3\%\) & \(10.6\%\) & \(13.8 / 11.1\%\) \\
      & \textbf{\texttt{GraphScent}} & \(70.6\%\) & \(58.8\%\) & \(28.9\%\) & \(41.2\%\) & \(44.6\%\) & \(64.9\%\) & \(38.9\%\) & \(10.6\%\) & \(13.8 / 11.1\%\) \\
    \midrule
    \multirow{10}{*}{\textbf{\texttt{12B}}}
      & \texttt{Flat BM25}          & \(69.2\%\) & \(63.0\%\) & \(35.5\%\) & \(39.3\%\) & \(48.3\%\) & \(63.5\%\) & \(37.9\%\) & \(9.6\%\)  & \(16.9 / 15.3\%\) \\
      & \texttt{Static Dir-PPR}     & \(70.1\%\) & \(64.0\%\) & \(32.7\%\) & \(39.8\%\) & \(47.9\%\) & \(62.6\%\) & \(36.0\%\) & \(10.6\%\) & \(16.9 / 13.9\%\) \\
      & \texttt{Undir-AST}          & \(71.6\%\) & \(65.4\%\) & \(35.1\%\) & \(41.7\%\) & \(47.4\%\) & \(65.9\%\) & \(40.3\%\) & \(17.0\%\) & \(20.0 / 20.8\%\) \\
      & \texttt{Fixed-Dir}          & \(70.1\%\) & \(64.9\%\) & \(35.1\%\) & \(44.1\%\) & \(47.9\%\) & \(65.4\%\) & \(41.2\%\) & \(20.2\%\) & \(20.0 / 16.7\%\) \\
      & \texttt{Ablate: Raw-Score}  & \(72.0\%\) & \(65.9\%\) & \(30.8\%\) & \(41.7\%\) & \(46.0\%\) & \(65.4\%\) & \(40.3\%\) & \(16.0\%\) & \(15.4 / 15.3\%\) \\
      & \texttt{Ablate: No-Gate}    & \(70.6\%\) & \(65.4\%\) & \(29.9\%\) & \(43.1\%\) & \(46.9\%\) & \(64.5\%\) & \(41.2\%\) & \(19.1\%\) & \(20.0 / 19.4\%\) \\
      & \texttt{Ablate: Rev-Dir}    & \(69.2\%\) & \(63.5\%\) & \(35.5\%\) & \(39.3\%\) & \(46.9\%\) & \(63.0\%\) & \(38.9\%\) & \(7.4\%\)  & \(15.4 / 16.7\%\) \\
      & \texttt{Ablate: No-Damp}    & \(71.1\%\) & \(63.5\%\) & \(30.8\%\) & \(40.8\%\) & \(44.1\%\) & \(66.4\%\) & \(42.7\%\) & \(13.8\%\) & \(13.8 / 13.9\%\) \\
      & \texttt{Ablate: No-2Hop}    & \(72.0\%\) & \(64.9\%\) & \(29.4\%\) & \(42.2\%\) & \(45.5\%\) & \(66.4\%\) & \(41.7\%\) & \(16.0\%\) & \(15.4 / 13.9\%\) \\
      & \textbf{\texttt{GraphScent}} & \(71.1\%\) & \(65.4\%\) & \(29.9\%\) & \(41.7\%\) & \(44.6\%\) & \(65.9\%\) & \(40.8\%\) & \(17.0\%\) & \(15.4 / 15.3\%\) \\
    \midrule
    \multirow{10}{*}{\textbf{\texttt{31B}}}
      & \texttt{Flat BM25}          & \(69.2\%\) & \(64.5\%\) & \(36.0\%\) & \(41.7\%\) & \(52.1\%\) & \(63.5\%\) & \(40.3\%\) & \(12.8\%\) & \(18.5 / 18.1\%\) \\
      & \texttt{Static Dir-PPR}     & \(70.1\%\) & \(64.9\%\) & \(35.1\%\) & \(42.2\%\) & \(52.6\%\) & \(62.6\%\) & \(40.8\%\) & \(13.8\%\) & \(18.5 / 18.1\%\) \\
      & \texttt{Undir-AST}          & \(74.9\%\) & \(68.2\%\) & \(36.0\%\) & \(41.7\%\) & \(48.8\%\) & \(69.7\%\) & \(44.1\%\) & \(17.0\%\) & \(15.4 / 16.7\%\) \\
      & \texttt{Fixed-Dir}          & \(74.4\%\) & \(67.8\%\) & \(35.5\%\) & \(43.6\%\) & \(48.8\%\) & \(68.7\%\) & \(44.1\%\) & \(20.2\%\) & \(20.0 / 19.4\%\) \\
      & \texttt{Ablate: Raw-Score}  & \(74.9\%\) & \(67.3\%\) & \(38.4\%\) & \(44.1\%\) & \(48.8\%\) & \(68.7\%\) & \(43.1\%\) & \(22.3\%\) & \(21.5 / 20.8\%\) \\
      & \texttt{Ablate: No-Gate}    & \(75.4\%\) & \(69.2\%\) & \(36.0\%\) & \(44.5\%\) & \(49.8\%\) & \(66.4\%\) & \(42.2\%\) & \(24.5\%\) & \(23.1 / 20.8\%\) \\
      & \texttt{Ablate: Rev-Dir}    & \(72.0\%\) & \(66.4\%\) & \(37.0\%\) & \(43.1\%\) & \(53.1\%\) & \(63.5\%\) & \(41.2\%\) & \(14.9\%\) & \(20.0 / 22.2\%\) \\
      & \texttt{Ablate: No-Damp}    & \(73.0\%\) & \(67.3\%\) & \(35.1\%\) & \(44.1\%\) & \(48.3\%\) & \(66.4\%\) & \(41.7\%\) & \(25.5\%\) & \(21.5 / 19.4\%\) \\
      & \texttt{Ablate: No-2Hop}    & \(75.4\%\) & \(68.2\%\) & \(36.0\%\) & \(44.5\%\) & \(49.8\%\) & \(66.8\%\) & \(43.6\%\) & \(25.5\%\) & \(26.2 / 23.6\%\) \\
      & \textbf{\texttt{GraphScent}} & \(74.9\%\) & \(68.7\%\) & \(36.0\%\) & \(43.6\%\) & \(48.3\%\) & \(66.8\%\) & \(42.7\%\) & \(23.4\%\) & \(21.5 / 19.4\%\) \\
    \bottomrule
  \end{tabular}}
\end{table}

\begin{table}[!htbp]
  \caption{Key paired bootstrap comparisons (\(2{,}000\) resamples) for \texttt{GraphScent} at \(B=24\) on \texttt{pooled\_211}, reporting exact task-level \texttt{Wins / Losses / Ties} and two-sided bootstrap \(p\)-values.}
  \label{tab:paired_cis}
  \centering
  \small
  \setlength{\tabcolsep}{3.0pt}
  \begin{tabular}{llcccc}
    \toprule
    \textbf{Scale} & \textbf{Paired Comparison (\(B=24\))} & \textbf{Subset (\(n\))} & \textbf{\(\Delta\) (pp) \([95\%\;\text{CI}]\)} & \textbf{W / L / T} & \textbf{\(p\)-val} \\
    \midrule
    \texttt{E4B} & \texttt{GraphScent LW} vs.\ \texttt{Undir-AST LW}   & All (\(211\))     & \(\mathbf{+5.7}\;[+0.5, +10.4]\) & \(22 / 10 / 179\) & \(0.037\) \\
    \texttt{E4B} & \texttt{GraphScent LW} vs.\ \texttt{Fixed-Dir LW}   & All (\(211\))     & \(+3.8\;[-0.9, +8.5]\)           & \(17 / 9 / 185\)  & \(0.126\) \\
    \texttt{E4B} & \texttt{GraphScent LW} vs.\ \texttt{Rev-Dir LW}     & All (\(211\))     & \(+4.3\;[+0.0, +8.5]\)           & \(16 / 7 / 188\)  & \(0.073\) \\
    \texttt{E4B} & \texttt{GraphScent LW} vs.\ \texttt{Flat BM25 LW}   & \texttt{BM25>10} (\(94\)) & \(\mathbf{+7.4}\;[+1.1, +13.8]\) & \(9 / 2 / 83\)    & \(0.033\) \\
    \texttt{E4B} & \texttt{GraphScent LW} vs.\ \texttt{GraphScent PW}  & All (\(211\))     & \(\mathbf{+12.3}\;[+6.2, +18.5]\) & \(37 / 11 / 163\) & \(<0.001\) \\
    \midrule
    \texttt{12B} & \texttt{GraphScent LW} vs.\ \texttt{Flat BM25 LW}   & \texttt{BM25>10} (\(94\)) & \(+7.4\;[-1.1, +16.0]\)          & \(13 / 6 / 75\)   & \(0.119\) \\
    \texttt{12B} & \texttt{GraphScent LW} vs.\ \texttt{Fixed-Dir LW}   & All (\(211\))     & \(-2.4\;[-6.6, +1.9]\)           & \(8 / 13 / 190\)  & \(0.346\) \\
    \texttt{12B} & \texttt{GraphScent LW} vs.\ \texttt{GraphScent PW}  & All (\(211\))     & \(\mathbf{+11.8}\;[+6.2, +17.5]\) & \(32 / 7 / 172\)  & \(<0.001\) \\
    \midrule
    \texttt{31B} & \texttt{GraphScent LW} vs.\ \texttt{Flat BM25 LW}   & \texttt{BM25>10} (\(94\)) & \(\mathbf{+10.6}\;[+2.1, +19.1]\) & \(14 / 4 / 76\)   & \(0.015\) \\
    \texttt{31B} & \texttt{GraphScent LW} vs.\ \texttt{Undir-AST LW}   & \texttt{BM25>10} (\(94\)) & \(+6.4\;[-2.1, +14.9]\)          & \(12 / 6 / 76\)   & \(0.182\) \\
    \texttt{31B} & \texttt{GraphScent LW} vs.\ \texttt{Fixed-Dir LW}   & \texttt{BM25>10} (\(94\)) & \(+3.2\;[-4.3, +10.6]\)          & \(8 / 5 / 81\)    & \(0.475\) \\
    \texttt{31B} & \texttt{GraphScent LW} vs.\ \texttt{GraphScent PW}  & All (\(211\))     & \(\mathbf{+7.6}\;[+2.4, +13.3]\)  & \(24 / 8 / 179\)  & \(0.004\) \\
    \bottomrule
  \end{tabular}
\end{table}

\begin{table}[!htbp]
  \caption{Per-repository \texttt{LW Commit@1} (\(B=24\)) across the four repositories in \texttt{public\_113} (\texttt{encode/httpx}, \(n=1\), is solved at \(100.0\%\) by every method across all three scales).}
  \label{tab:per_repo}
  \centering
  \small
  \setlength{\tabcolsep}{5.0pt}
  \begin{tabular}{llccccc}
    \toprule
    \textbf{Repository (\(n\))} & \textbf{Scale} & \texttt{Flat BM25} & \texttt{BAGEL-GP} & \texttt{Undir-AST} & \texttt{Fixed-Dir} & \textbf{\texttt{GraphScent}} \\
    \midrule
    \multirow{3}{*}{\texttt{fastapi/fastapi} (\(56\))}
      & \texttt{E4B} & \(35.7\%\) & \(32.1\%\) & \(33.9\%\) & \(30.4\%\) & \(\mathbf{35.7\%}\) \\
      & \texttt{12B} & \(30.4\%\) & \(33.9\%\) & \(35.7\%\) & \(35.7\%\) & \(\mathbf{37.5\%}\) \\
      & \texttt{31B} & \(32.1\%\) & \(32.1\%\) & \(33.9\%\) & \(33.9\%\) & \(\mathbf{37.5\%}\) \\
    \midrule
    \multirow{3}{*}{\texttt{Textualize/rich} (\(44\))}
      & \texttt{E4B} & \(\mathbf{40.9\%}\) & \(31.8\%\) & \(27.3\%\) & \(36.4\%\) & \(38.6\%\) \\
      & \texttt{12B} & \(40.9\%\) & \(38.6\%\) & \(38.6\%\) & \(\mathbf{47.7\%}\) & \(29.5\%\) \\
      & \texttt{31B} & \(40.9\%\) & \(38.6\%\) & \(40.9\%\) & \(47.7\%\) & \(\mathbf{50.0\%}\) \\
    \midrule
    \multirow{3}{*}{\texttt{psf/requests} (\(12\))}
      & \texttt{E4B} & \(50.0\%\) & \(50.0\%\) & \(50.0\%\) & \(\mathbf{58.3\%}\) & \(\mathbf{58.3\%}\) \\
      & \texttt{12B} & \(50.0\%\) & \(50.0\%\) & \(\mathbf{58.3\%}\) & \(\mathbf{58.3\%}\) & \(\mathbf{58.3\%}\) \\
      & \texttt{31B} & \(58.3\%\) & \(\mathbf{66.7\%}\) & \(58.3\%\) & \(58.3\%\) & \(58.3\%\) \\
    \bottomrule
  \end{tabular}
\end{table}

\FloatBarrier
\section{Negative Results and Design Explorations}
\label{app:negative}

Several plausible hypotheses failed during our investigation; reporting them clarifies why GraphScent adopts frozen \(T=0\) (\texttt{thinking=off}) scoring over typed AST relations.

\paragraph{Supervised LoRA fine-tuning on commit patches suppresses neighbor scent.}
We trained a rank-16 LoRA adapter on \texttt{Gemma 4 31B} using historical issue-to-patch relevance labels (\(S=3\) strictly for edited functions and \(S=0\) for unedited functions). Evaluating \(1{,}540\) nodes on \texttt{public\_113} (Table~\ref{tab:legacy}), LoRA slightly lowered direct \texttt{gold} vs.\ \texttt{far} \(\mathrm{AUROC}\) (\(0.713 \to 0.704\)) and \emph{significantly degraded} 1-hop structural neighbor scent (\(\mathrm{AUROC}(\texttt{near}\;\text{vs.}\;\texttt{far})\) fell from \(0.570\;[.53,.61]\) to \(0.536\;[.51,.56]\), and embedding-neighbor scent fell from \(0.497\) to \(0.465\)). Training binary target classifiers teaches the LLM to suppress high scores on unedited callers, eroding the proximal cue required for multi-hop navigation.

\paragraph{Chain-of-thought (\texttt{thinking=on}) vs.\ direct scoring (\texttt{thinking=off}).}
Enabling \texttt{thinking=on} on \texttt{31B} modestly increased direct \texttt{gold} vs.\ \texttt{far} \(\mathrm{AUROC}\) (\(0.713 \to 0.731\)), left 1-hop structural neighbor scent unchanged (\(0.570 \to 0.567\)), lowered strict JSON parse completion from \(100.0\%\) (\(1540/1540\)) to \(96.0\%\) (\(1479/1540\)) under bounded output caps, and increased output-token latency by \(>6\times\). Direct \(T=0\) scoring (\texttt{thinking=off}) is therefore strictly preferable for inner-loop node probing.

\paragraph{Historical git \texttt{cochange} and supplied 256-d embeddings.}
Neither mining commit history for frequently co-edited entities (\texttt{cochange}: \(\mathrm{Scent}=0.463\text{--}0.498\)) nor propagating Gaussian-process beliefs over the benchmark's 256-d embeddings (\texttt{knear\_comp256}: \(\mathrm{Scent}=0.436\text{--}0.470\); \texttt{BAGEL-GP} \(38.9\%\) vs.\ \(45.1\%\) \texttt{GraphScent} at \texttt{31B}) improved over \texttt{Flat BM25} (\(38.9\%\)).

\begin{table}[!htbp]
  \caption{Controlled comparison on \(1{,}540\) \texttt{public\_113} nodes across frozen \texttt{Gemma 4 31B} (\texttt{thinking=off}), \texttt{thinking=on}, and rank-16 commit-patch LoRA fine-tuning.}
  \label{tab:legacy}
  \centering
  \small
  \setlength{\tabcolsep}{4.8pt}
  \begin{tabular}{lcccc}
    \toprule
    \textbf{31B Judge Variant} & \textbf{JSON Parse} & \textbf{\texttt{gold} vs.\ \texttt{far}} & \textbf{1-Hop AST vs.\ \texttt{far}} & \textbf{\texttt{knear} vs.\ \texttt{far}} \\
    \midrule
    Frozen \texttt{31B} (\texttt{thinking=off}) & \(100.0\%\) & \(0.713\;[.66,.76]\) & \(\mathbf{0.570}\;[.53,.61]\) & \(0.497\;[.46,.53]\) \\
    Frozen \texttt{31B} (\texttt{thinking=on})  & \(96.0\%\)  & \(\mathbf{0.731}\;[.68,.78]\) & \(0.567\;[.53,.61]\) & \(0.522\;[.48,.56]\) \\
    Patch-LoRA (\texttt{thinking=off})          & \(100.0\%\) & \(0.704\;[.66,.75]\) & \(0.536\;[.51,.56]\) & \(0.465\;[.44,.49]\) \\
    \bottomrule
  \end{tabular}
\end{table}

\FloatBarrier
\newpage
\section{Token Accounting, Hardware Footprint, and Prompt Templates}
\label{app:prompts}

\begin{table}[!htbp]
  \caption{Exact prompt-token distributions and single-GPU VRAM footprints across Gemma~4 scales (\(B=24\), \(25\) stateless calls per task, \texttt{max\_tokens=32} output cap per call).}
  \label{tab:tokens}
  \centering
  \small
  \setlength{\tabcolsep}{3.3pt}
  \begin{tabular}{lccccc}
    \toprule
    \textbf{Model (\texttt{w4a16})} & \textbf{VRAM} & \textbf{Node Med / P90} & \textbf{LW Med / P90} & \textbf{Cum.\ In (Med / Mean / P90)} & \textbf{Cum.\ Out} \\
    \midrule
    Gemma 4 \texttt{E4B} & \(10.42\,\text{GiB}\) & \(657 / 1{,}266\) & \(2{,}197 / 3{,}142\) & \(18{,}886 / 22{,}404 / 35{,}603\) & \(\sim 225\) \\
    Gemma 4 \texttt{12B} & \(8.51\,\text{GiB}\)  & \(669 / 1{,}295\) & \(2{,}116 / 3{,}086\) & \(18{,}309 / 21{,}682 / 35{,}119\) & \(\sim 225\) \\
    Gemma 4 \texttt{31B} & \(20.28\,\text{GiB}\) & \(691 / 1{,}323\) & \(2{,}050 / 3{,}147\) & \(15{,}211 / 18{,}162 / 32{,}867\) & \(\sim 225\) \\
    \bottomrule
  \end{tabular}
\end{table}
\FloatBarrier

\paragraph{Verbatim stateless node-scoring prompt (\(T=0\), \texttt{thinking=off}, \texttt{max\_tokens=32}).}
\begin{small}
\begin{verbatim}
You are a software bug localization judge. Given a GitHub issue and a folded view
of ONE Python entity from the repository, rate bug relevance on {0,1,2,3}:
  3 = Root-cause defect requiring modification is directly in this node.
  2 = Strongly involved or an immediate caller/enclosing scope of the defect.
  1 = On the broader execution path or loosely related module, unlikely edited.
  0 = Unrelated to the issue.
Respond ONLY with JSON: {"score": <0|1|2|3>}
\end{verbatim}
\end{small}

\paragraph{Verbatim top-6 listwise commitment prompt (\(T=0\), \texttt{thinking=off}, \texttt{max\_tokens=32}).}
\begin{small}
\begin{verbatim}
You are selecting the single primary code entity that must be edited to resolve the
GitHub issue below. Six candidate entities [0..5] found by repo search
are shown with their folded code views. Respond ONLY with JSON: {"best_idx": <0..5>}
\end{verbatim}
\end{small}

\end{document}